\documentclass[aps, reprint, a4paper, pra, showpacs]{revtex4-1}
\usepackage{amsmath}
\usepackage{graphicx}
\usepackage{epstopdf}
\usepackage{subcaption}
\usepackage{bm}

\newcommand{\beq}{\begin{equation}}
\newcommand{\eeq}{\end{equation}}
\newcommand{\ali}[1]{\begin{align} #1 \end{align}}
\newcommand{\spl}[1]{\begin{split} #1 \end{split}}
\newcommand{\ve}{\varepsilon}
\newcommand{\updown}{\uparrow \! \downarrow}
\newcommand{\Li}[1]{\mathrm{Li}_{#1}}
\newcommand{\order}[1]{\mathcal{O} \! \left( #1 \right)}
\newcommand{\N}{\mathcal{N}_c}
\newcommand{\intd}{\mathrm{d}}

\begin{document}

\title{Non-Analytic Crossover Behavior of SU($\mathcal{N}_c$) Fermi Liquid}
\date{\today}
\author{Pye Ton How}
\affiliation{Institute of Physics, Academia Sinica, Taipei 115,
	Taiwan}

\author{Sung-Kit Yip}
\affiliation{Institute of Physics, Academia Sinica, Taipei 115,
	Taiwan}
\affiliation{Institute of Atomic and Molecular Sciences, Academia
	Sinica, Taipei 106, Taiwan}

\begin{abstract}
We consider the thermodynamic potential of a dilute Fermi gas with a
contact interaction, at both finite temperature $T$ and non-zero
effective magnetic fields $\mathbf{H}$, and derive the equation of
state analytically using second order perturbation theory.  Special
attention is paid to the non-analytic dependence of $\Omega$ on
temperature $T$ and (effective) magnetic field $\mathbf{H}$, which
exhibits a crossover behavior as the ratio of the two is continuously
varied.  This non-analyticity is due to the particle-hole pair
excitation being always gapless and long-ranged.  The non-analytic
crossover found in this paper can therefore be understood as an analog
of the Ginzberg-Landau critical scaling, albeit only at the
sub-leading order.  We extend our results to an $\mathcal{N}_c$-
component Fermi gas with an $\mathrm{SU}(\mathcal{N}_c)$-symmetric
interaction, and point out possible enhancement of the crossover
behavior by a large $\mathcal{N}_c$.
\end{abstract}


\maketitle

\section{Introduction}

The Fermi liquid (FL) paradigm is an important cornerstone of our
understanding of nature.  It was originally conceived as a
phenomenological theory for liquid Helium-3, but turned out to be
generally a good description for most physical systems with Fermionic
degrees of freedom at low enough temprature.  There has long been a
consensus \cite{Doniach1966, Amit1968, Mota1969, Pethick1973, 
Carneiro1975, Carneiro1977, Belitz1997, Chitov2001, Misawa2001, 
Chubukov2003, Betouras2005, Chubukov2005, Chubukov2006, Maslov2009}
that the thermodynamic behavior of an FL must be non-analytic, in
contrast to the Ginzburg-Landau (GL) theory which assumes that the
free energy takes an analytic form away from a phase transition.
Historically, the specific heat of normal ${}^3\text{He}$ was the
earliest experimentally studied example \cite{Abel1966, Greywall1983},
where the observed trend cannot be fitted to an analytic function. 
Theoretical efforts \cite{Doniach1966, Amit1968, Pethick1973,
Carneiro1975, Chubukov2003, Chubukov2006} indicate, to leading order,
a $T^3 \ln T$ correction on top of the linear $T$ dependence from the
leading-order FL behavior.  This non-analytic correction is a generic 
feature of any FL, in the sense that it is entirely captured by 
considering the interaction and scattering between Landau
quasi-particles on the Fermi surface.  This term has also been
studied in the context of heavy fermion metals \cite{Coffey1986, 
VanDerMeulen1990}.  In ordinary metal, however, it was concluded 
\cite{Coffey1988} that the effect will be too small to be
experimentally observed.

In the context of an electron liquid, a magnetic field causes Zeeman
split between the two spin components.  It was later realized that
the magnetic response of an electron liquid is also non-analytic
beyond the leading order\cite{Belitz1997, Chubukov2003, Betouras2005,
Maslov2009}, with the underlying physics closely related to the
temperature case.  Theories indicate a $H^2 \ln H$ correction to
the constant Pauli susceptibility.  In two space dimensions, similar
considerations lead to the prediction of a $T^2$ correction to
specific heat and a $\vert H \vert$ correction to spin susceptibility
\cite{Belitz1997, Chitov2001, Chubukov2003, Betouras2005, Maslov2009}.

This non-analytic magnetic response has a much more dramatic
consequence: it can change the order of the itinerant Ferromagnetic
quantum critical point \cite{Vojta1997, Belitz1999, Maslov2009} from
second order, as dictated by the GL paradigm, to weakly first order
\cite{Pfleiderer1997}.

The particle-hole pair excitation around the Fermi surface has been
identified as the cause of this non-analytic behavior
\cite{Doniach1966, Amit1968a, Pethick1973, Chubukov2006}.  Such a
pair is always gapless in the normal phase.  The infrared singularity
of the pair's Green's function, while not strong enough to cause a
full-fledged divergence, results in the non-analyticity.

Yet it remains difficult to draw a more precise conclusion beyond
the statement that theories and experiments agree qualitatively.  
Theoretically, even within the Fermi liquid picture, the calculations
were usually performed by considering only a subset of all possible 
interaction processes \cite{Doniach1966, Amit1968a, Pethick1973, 
Chubukov2003}, where the omitted processes solely gives rise to
analytic terms.  One then obtains the non-analytic term, but on top
of an unknown background of analytic contributions.  Experimentally,
even for the well-studied case of ${}^3\text{He}$ specific heat, the
uncertainty in interacting parameters is large enough \cite{Dy1969, Baym1991} to prevent a more meaningful comparison (see, for example,
the discussion of \cite{Pethick1973, Chubukov2006}.)  To the best of
out knowledge, the $H^2 \ln H$ behavior of spin susceptibility has
not been observed.  However there are experimental evidences of
its two-dimensional counterpart: ref \cite{Zhang2009} pointed out
that the normal state of iron pnictide exhibits a spin susceptibility
that increases linearly with temperature \cite{Wang2009,
Klingeler2010}.  This is consistent with the linear non-analyticity
in two dimensions when temperature is dominant, as discussed in
\cite{Korshunov2009}.

In this paper, we theoretically study the non-analytic effects in
the context of a dilute Fermi gas in three space dimensions (3D),
in second order perturbation theory.  This choice of theoretical
model is made with possible cold atom experiments in mind.

Experimentally, cold quantum gas has several advantages over other
realizations of FL.  First of all, the interaction between particles
is well approximated by contact interaction, and is highly tunable
through the Feshbach resonance technique (\cite{Inouye1998}, and see
\cite{Chin2010} for a review).  One may realize the
weakly-interacting dilute limit, amenable to a perturbative treatment.
Secondly, through the use of a non-homogeneous trap and the local
density approximation, one has direct access to the equation of state
of the gas \cite{Cheng2007, Navon2010, Ku2012, VanHoucke2012,
Desbuquois2014}.  Finally, one is not confined to two-component
spin-$\frac{1}{2}$ Fermions: isotopes such as ${}^{173}\text{Yb}$
and ${}^{87}\text{Sr}$ have large pseudo-spins \cite{Fukuhara2007,
Cazalilla2009, DeSalvo2010, Cazalilla2014}, which enhance the effects
of interaction, and potentially make the non-analytic part more
visible experimentally.

At the center of our attention is the thermodynamic potential
$\Omega$ of the gas.  Extending from our previous paper
\cite{Cheng2017} (referred to as the prequel hereafter), we shall
study the behavior of $\Omega$ using perturbation theory to second
order, focusing on the interplay of temperature and magnetic field.
In particular, we investigated the crossover between the zero-magnetic
field limit and the zero-temperature limit.  We obtain an equation
of state for the gas, quantifying both the analytic and the
non-analytic contributions to $\Omega$, thereby facilitating a
direct comparison with future experiments.

Since the particle-hole pair excitation is always gapless, it is
legitimate to ask if the resultant physics shares any similarities
with the usual GL critical phenomeology.  We will see that the
crossover behavior is strongly reminiscent of a quantum critical
point, albeit only at the sub-leading order.  One may claim that
a non-magnetic Fermi liquid is, in a sense, always ``critical'',
regardless of the interaction strength.  A similar idea was explored
by Belitz, Kirkpatrick and Vojta \cite{Belitz2005}.

\section{Thermodynamic Potential}

\subsection{Model Hamiltonian}

The $\N$-component fermion gas is modeled with anticommuting
quantum fields $\psi_a$, where the index $a$ runs from 1 to $\N$.
The generalized Zeeman shifts in an $\mathrm{SU}(\N)$-invariant
theory is given by a traceless Hermitian matrix $\mathbf{H}$, but
without loss of generality it may be put into a diagonal form with
an appropriate $\mathrm{SU}(\N)$ transformation.  We therefore write
the free Hamiltonian as the following:
\beq
H_0 =
	\sum_{a=1}^{\N} \int \! \mathrm{d}^3 x \, \psi^{\dag}_{a}
		\left( - \frac{\nabla^2}{2m} - \mu_0
			- H_a
		\right) \psi_{a}.
\label{freeHamiltonian}
\eeq
Here $H_a$ is an eigenvalue of $\mathbf{H}$.  The traceless condition
of $\mathbf{H}$ translates to $\sum_a H_a = 0$.  It is sometimes also 
convenient to consider the species-dependent effective chemical 
potential:
\beq
\mu_a \equiv \mu_0 + H_a.
\eeq

We also define the associated momentum scale
$k_{\mu} = \sqrt{2 m \mu_0}$ and the analogy of Fermi velocity
$v_{\mu} = k_{\mu}/m$.  To the order of approximation adopted in
this paper, these are interchangeable with the actual Fermi momentum
and velocity $k_F$ and $v_F$.

In this paper, we shall treat $\mu_0$ and $\mathbf{H}$, rather than
the fermion number density, as the ``tuning knobs'' of the system.
Experimentally, the (position-dependent) chemical potential  of a
trapped quantum gas within the local density approximation is
readily available.  So we do not see this as a difficulty.

We employ a zero-ranged interaction:
\beq
H_I = \frac{4 \pi a}{m} \; \sum_{a,b = 1}^{\N}
		\int \! \mathrm{d}^3 x \,
			\psi^{\dag}_{a}(x) \psi^{\dag}_{b}(x)
			\psi_{b}(x) \psi_{a}(x).
\label{interaction}
\eeq
The quantity $a$ is the scattering length of the zero-ranged
two-body potential.  We shall perform our calculation in the dilute
limit, where $(a \, k_F)$ is a small expansion parameter.  We work
in the units $\hbar = k_B = 1$.

Two-particle scattering amplitudes of this model diverges in the
Cooper channel.  This is the usual pathology of a $\delta$-function
potential, and can be absorbed by renormalization.  In the following,
we shall implicitly assume that all such divergences are removed.
Related to this diverging behavior is a pairing instability in the
Cooper channel at an exponentially small transition temperature.
This instability shall be ignored in all subsequent discussions.

Staring from this point, we will consider the case $\N = 2$, as the
crossover advertised in the beginning is essentially an effect
between two spin components.  The generalization to a generic
$\N > 2$ will be discussed later in the paper.

For the $\N = 2$ case, we denote the two spins $\uparrow$ and
$\downarrow$, with the convention $H_{\uparrow} = H/2$ and
$H_{\downarrow} = -H/2$.  Without loss of generality one can always
assume $H \geq 0$.

\subsection{The Origin of Non-Analyticity}

\label{origin}

For an in-depth discussion of the result quoted in this subsection,
we refer our readers to the work of Chubukov, Maslov and Millis 
\cite{Chubukov2006}, and the references there in.

\begin{figure}
	\begin{subfigure}{0.4 \textwidth}
	\includegraphics[width=0.6 \textwidth]{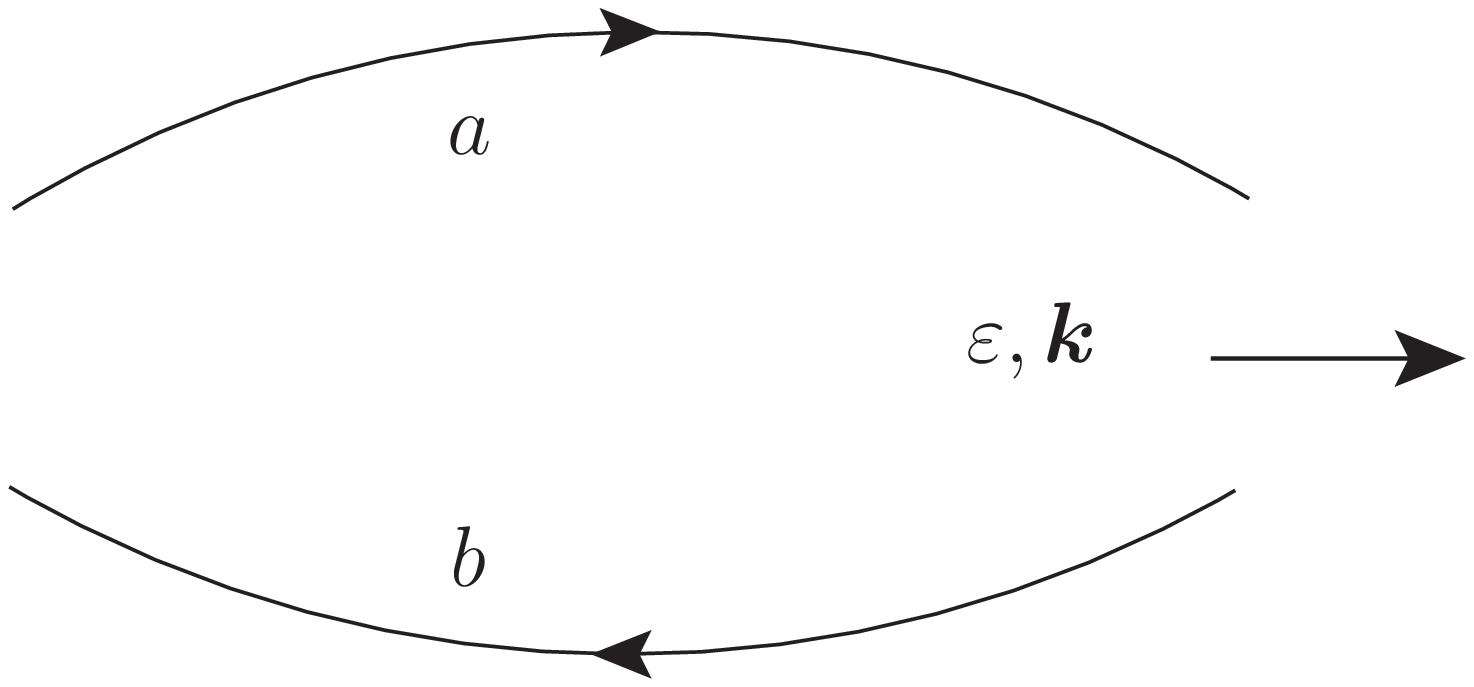}
	\caption{\label{lindhard} Particle-hole pair}
	\end{subfigure}

	\begin{subfigure}[b]{0.2\textwidth}
	\includegraphics[width = 0.7 \textwidth]{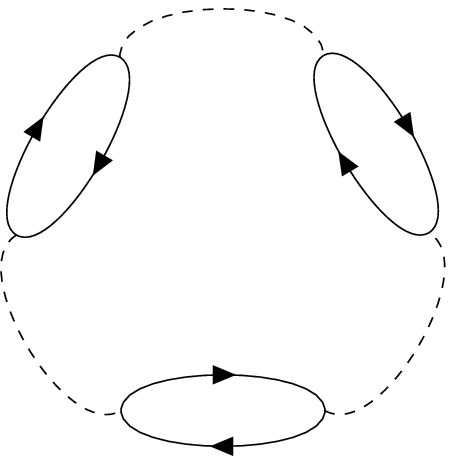}
	\caption{\label{ring3} Ring diagram}
	\end{subfigure}
	\begin{subfigure}[b]{0.2\textwidth}
	\includegraphics[width = 0.7 \textwidth]{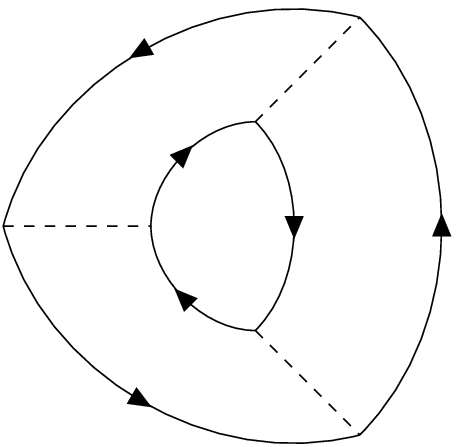}
	\caption{\label{ladder3} Ladder diagram}
	\end{subfigure}
	\caption{Figure (a) represents the particle-hole bubble
	$\Pi_{ab}(\ve, \bm{k})$. Note that the spins $a$ and $b$
	can be different.  (b) and (c) are examples of ring and
	ladder diagrams respectively, each showing three particle-hole
	pairs joined together.}
\end{figure}

For the $\N = 2$ case, the specific heat and spin susceptibility
have been shown to receive logarithmic corrections, and the
particle-hole pair excitation is solely responsible for non-analytic
behavior of $\Omega$.  As shown in figure \ref{lindhard}, we denote
the spins of the particle and hole in such a pair as $a$ and $b$ 
respectively, which may or may not be the same.  We denote the
Green's function for such pair excitation $\Pi_{ab}$.

The thermodynamic potential $\Omega$ can be computed by summing over
vacuum Feynman diagrams.  The pair excitation modes contribute to
$\Omega$ via two classes of diagrams: the ring diagrams where each
bubble consists of the same spin, and the ladder diagrams where the
particle and hole legs have different spins.  Figures \ref{ring3}
and \ref{ladder3} are examples with three particle-hole pairs.

These pair fluctuation are bosonic and remain soft down to zero
momentum.  The long-ranged correlation of these bosonic modes is
not strong enough to cause full-fledged infrared divergence in the
present case, but results in weaker logarithmic corrections only
at higher orders.  This is the origin of the non-analyticity.  For
a review of this ``soft mode'' paradigm, and in particular how it
affects the critical behavior, see \cite{Belitz2005} and the
references therein.

These soft modes must be cut off by some relevant infrared scale.
Temperature is an obvious candidate.  In the presence of
$(H_a - H_b) \neq 0$, it can be seen that the energy of pair
excitation $\Pi_{ab}$ is shifted by $(H_a - H_b)$;  therefore the
magnetic field can also serve as the cutoff.  One expects the larger
of the two scales to dominate, and this hints at possible crossover 
behavior when the ratio $T/H$ is varied continuously between the two 
extremes \cite{Betouras2005}.  However, each particle-hole pair in
the ring diagram is of the same spin, and is insensitive to the
magnetic field.  The crossover behavior is thus exhibited only in
the ladder-type non-analyticity.

\begin{figure}
	\begin{subfigure}[b]{0.2\textwidth}
	\includegraphics[width = 0.8 \textwidth]{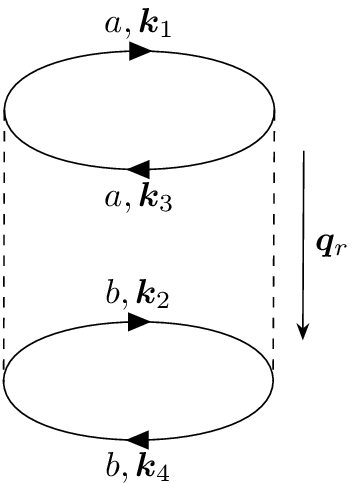}
	\caption{\label{ring2} Ring diagram}
	\end{subfigure}
	\begin{subfigure}[b]{0.2\textwidth}
	\includegraphics[width = \textwidth]{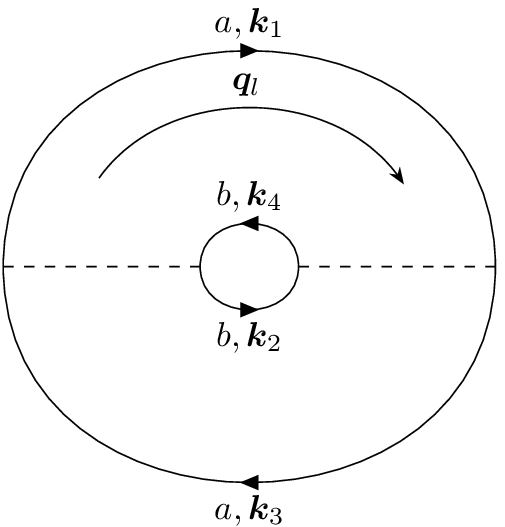}
	\caption{\label{ladder2} Ladder diagram}
	\end{subfigure}
	\caption{\label{omega2a} The ring and ladder forms of the
	same second-order vacuum Feynman diagram.  The small-momentum
	limits of the forms corresponds to the limit where
	$\bm{q}_{r} = \bm{k}_1 - \bm{k}_3$ or
	$\bm{q}_{l} = \bm{k}_1 - \bm{k}_4$ vanishes, respectively.
	This term is labeled as $\Omega_{2a}$.}
\end{figure}

In this paper we work to second order in perturbation theory.  The
ring and ladder diagram at second order is actually one and the same,
as shown in figure \ref{omega2a}.  However, the two small-momentum
limits refer to distinct regions of the momentum integral.

At second order, the ring diagram is known to yield further
non-analytic terms in the region $q_r \approx 2 k_F$
\cite{Chubukov2006}.  Historically this has been linked to the
dynamic Kohn anomaly \cite{Amit1968, Pethick1973, Chubukov2006}.  
However, this non-analytic contribution only comes from the limit
where $\bm{q}_1$ and $\bm{q}_2$ are anti-parallel \cite{Amit1968},
and can be identified with the small-$q_l$ limit of the ladder
diagram \cite{Chubukov2006, Maslov2009}.  Conversely, the ladder
diagram also contributes to the non-analyticity around
$q_l \approx 2k_F$, and this translates to the small-$q_r$ limit of
the ring diagram.

We argue that, rather than the traditional ``zero-and-$2k_F$''
picture, it is more natural to look at the non-analyticity as coming
solely from the zero-momentum limit of particle-hole pairs, but
then consider all possible spin combinations.

\subsection{Possible Form of $\Omega$}

The thermodynamic potential is $\Omega = -T \ln Z$.  In the
thermodynamic limit, it is more convenient to consider the
intensive quantity $\Omega/V$, where $V$ is the volume of space.

The usual GL paradigm dictates that $\Omega$ be an analytic function
of $T$ and $H$. Coupled with the symmetry of the problem, one expects
that $\Omega$ can be expanded as a polynomial of $T^2$ and $H^2$
only.  But the known $T^3 \ln T$ specific heat and $H^2 \ln H$ spin
susceptibility implies that the GL picture is no good already at
fourth order.

Define the dimensionless quantities $t = T/\mu_0$ and $h = H/\mu_0$.
On dimensional ground, and with the knowledge that the Sommerfeld
expansion cannot generate odd powers of $T$, one writes down 
schematically the possible form of $\Omega$, omitting all
coefficients:
\beq
\frac{\Omega}{V} \sim 
	v_{\mu} k_{\mu}^4 \left\lbrace
		1 + \left(t^2 + h^2 \right) + F_4(t, h)
		+ \dots
	 \right\rbrace.
\label{omegaDimAnal}
\eeq

Here the (generalized) fourth order term $F_4$ is defined to be the
sum of all terms that scale as $(\text{energy})^4$, up to possible
logarithmic dependence.  We know that $F_4$ must be non-analytic at
$T = H = 0$.  Its behavior near the origin of the $(T, H)$-plane will
depend on the direction of approach.  In particular, if one attempts
a double expansion of $F_4$ in $t$ and $h$, the result will depend
on the order in which the expansions are carried out.

\subsection{What Is Fermi Liquid and What Is Not?}

First verified by Pethick and Carneiro\cite{Pethick1973}, an
oft-repeated observation is that the leading logarithmic correction
is a ``universal'' feature of any FL.  This raises the question: in
the expression \eqref{omegaDimAnal}, what can be considered
``universal''?

We try to address this question in the context of a dilute Fermi gas.
FL is then a low-energy effective theory of the model, with only
degrees of freedom near the Fermi surface.  One can take a linearized
quasiparticle dispersion and an approximated constant density of state
around the Fermi surface as the working definition of this effective 
theory.  All the higher-energy modes are integrated out, renormalizing
the parameters of this effective theory.

Furthermore, FL is only accurate when all external scales in the
problem, such as $T$ and $H_a$, are dwarfed by the Fermi sea.  In
other words, $\mu_0$ and $k_F$ should be considered essentially
infinite when compared with other scales. This means that, at high
enough order, terms in \eqref{omegaDimAnal} will eventually be deemed
infinitesimal and outside the scope of FL.

For the non-interacting gas, one can calculate $\Omega$ exactly.  It
can be shown that the leading correction to the free gas (second order
in $t$ and $h$) in \eqref{omegaDimAnal} depends only on Fermi velocity
and density of state at Fermi surface \footnote{For the general case
of $\N > 2$, third order terms in magnetic field is possible; see
section \ref{largeN}.  But it is also determined exclusively by FL
parameters.}.  So one can conclude that these are within the FL
picture, while fourth and higher order terms are beyond FL.  In
contrast, the non-analytic terms at fourth order do come from the
Fermi surface only, as mentioned above.

To go beyond dilute limit and perturbation theory, one can replace
Fermi velocity, density of state, and scattering amplitudes of 
quasiparticles with their fully renormalized values, as suggested in
refs \cite{Chubukov2006, Maslov2009}.  By construction, this simple
replacement yields FL results (second and third order terms, and the
fourth order logarithmic correction) that remain valid in the
strongly-interacting regime.  This however will not apply to terms
outside the scope of FL, and we lose all ability to calculate them
in the strongly interacting regime.

\section{Sommerfeld Expansion of $\Omega$ for Fermi Gas}

In this section, we take a break from the FL picture, and attempt to
evaluate the thermodynamic potential of a weakly interacting Fermi
gas.  This will confirm some assertions made in section \ref{origin},
and also give us some hints on the possible form of $F_4$.

\begin{figure}
	\begin{subfigure}[b]{0.4\textwidth}
		\includegraphics[width = 0.6\textwidth]{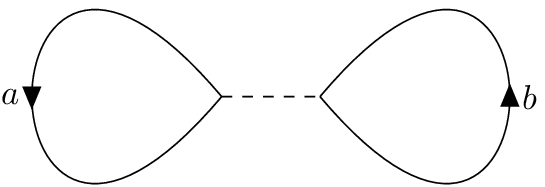}
		\caption{\label{omega1} $\Omega_{1}$}
	\end{subfigure}

	\begin{subfigure}[b]{0.4\textwidth}
		\includegraphics[width = 0.9\textwidth]{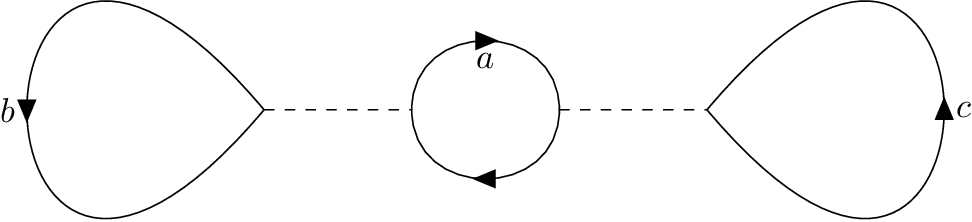}
		\caption{$\Omega_{2b}$}
	\end{subfigure}
	\caption{\label{HFdiagrams} The Hartree-Fock diagrams
	contributing to the thermodynamic potential $\Omega$ to
	second order.}
\end{figure}

We wish to calculate the thermodynamic potential of the gas.  To
second order of perturbation theory, there are three Feynman diagrams
to be included.  Following the notation of the prequel
\cite{Cheng2017}, we label these three terms as $\Omega_{1}$,
$\Omega_{2a}$ and $\Omega_{2b}$, respectively.  Depicted in Figure
\ref{HFdiagrams}, $\Omega_1$ and $\Omega_{2b}$ are part of the
Hartree-Fock approximation and are analytic in $T$ and $H$.  On the
other hand, $\Omega_{2a}$ as shown in Figure \ref{omega2a} is both
a ring and a ladder diagram, and is solely responsible to
non-analyticity of $\Omega$ at this level of approximation.

In this section, we will first attempt an expansion in $T$, writing
$\Omega = \alpha_0(H) + \alpha_2(H)\,  t^2 + \dots$ at finite $H$.
Analytic closed form solutions of these $H$-dependent coefficients
can be obtained, but we found that it is much more elucidating to
further expand each coefficient in series of $H$.  Please see
appendix \ref{sommerfeldAppen} for more detail.

To maintain consistency with the prequel, we define the dimensionless
lowercase $\omega$'s via
\beq
\spl{
\frac{\Omega}{V} = 
	\frac{k_{\mu}^3}{6\pi^2} \frac{k_{\mu}^2}{2 m}
	\sum_{a = 1}^{\N} \Bigg [
		\omega_0^{(a)}
		+ (k_{\mu} a) \sum_{b \neq a} \omega_1^{(ab)} \\
		+ (k_{\mu} a)^2 \left(
			\sum_{b \neq a} \omega_{2a}^{(ab)}
			+ \sum_{b \neq a} \sum_{c \neq a}
				\omega_{2b}^{(abc)}\right)
			+ \dots
	\Bigg ].
}
\label{dimless}
\eeq
where each $\omega_x$ originates from the respective $\Omega_{x}$
with the same label.  We have temporarily restored $\N$ in the above
expression.  For $\N = 2$, the sum over spin is quite trivial and we
define the spin-symmetrized version:
\begin{subequations}
\ali{
w_0 =& \;
	\frac{1}{2} \left(\omega_{0}^{(\uparrow)}
	+ \omega_{0}^{(\downarrow)} \right) ; \\
w_1 =& \;
	\frac{1}{2}\left(
		\omega_{1}^{(\uparrow \downarrow)}
		+ \omega_{1}^{(\downarrow\uparrow)}
	\right) ; \\
w_{2a} =& \;
	\frac{1}{2}\left(
		\omega_{2a}^{(\uparrow \downarrow)}
		+ \omega_{2a}^{(\downarrow \uparrow)}
	\right) ; \\
w_{2b} =& \;
	\frac{1}{2}\left(
		\omega_{2b}^{(\uparrow \downarrow \uparrow)}
		+ \omega_{2b}^{(\downarrow \uparrow \downarrow)}
	\right) .
}
\label{dimlessN2}
\end{subequations}

\subsection{Free Gas and Hartree-Fock Contributions}

Let $\epsilon(k)$ denotes the kinetic energy of free gas, $n_a(x)$
the Fermi function for fermions with spin $a$, and $N_a^0$ the
non-interacting number density for these fermions:
\begin{subequations}
\ali{
\epsilon(k) \equiv& \; \frac{k^2}{2m} ; \\
n_a(k) \equiv& \; \left(
	e^{\beta(\epsilon(k)-\mu_a)} + 1
	\right)^{-1} ; \\
N_a^0 \equiv& \int \! \! \frac{\intd^3 k}{(2 \pi)^3} \, 
	n_a \! (k).
}
\label{freeGasNotation}
\end{subequations}

The thermodynamic potential of a free gas is given by
\beq
\frac{\Omega_0}{V} =
	T \sum_{a = \uparrow, \downarrow}
	\int \!\! \frac{\intd^3 k}{(2 \pi)^3} \,
		\ln \!\left(
			1 - n_a \! \left(k\right)
			\right).
\eeq

Likewise, the two Hartree-Fock terms $\Omega_1$ and $\Omega_{2b}$
are:
\ali{
\frac{\Omega_1}{V} =& \left(\frac{4 \pi a}{m}\right)
	N_{\uparrow}^{0} N_{\downarrow}^{0}; \\
\frac{\Omega_{2b}}{V} =&
	\left(\frac{4 \pi a}{m}\right)^{\! 2}
	\frac{1}{2} \left[
		\left(
			\frac{\partial N_{\uparrow}^0}{\partial \mu_0}
		\right)
		N_{\downarrow}^0 N_{\downarrow}^0
		+ \left(
			\frac{\partial N_{\downarrow}^0}{\partial \mu_0}
		\right)
		N_{\uparrow}^0 N_{\uparrow}^0
	\right] .
}

From here one can identify the associated dimensionless $\omega_{0}$,
$\omega_{1}$ and $\omega_{2b}$.  Up to fourth order in $t$ and $h$,
they are
\begin{widetext}
\begin{subequations}
\ali{
w_0 = \;&
	-\frac{2}{5} - \frac{\pi^2}{4} t^2
	-\frac{1}{16} h^2 + \frac{7 \pi^4}{960} t^4 +
	\frac{1}{1024} h^4 +\frac{\pi^2}{128} t^2 h^2 ;\\
w_1 = \;&
	\frac{2}{3\pi} + \frac{\pi}{6} t^2 - \frac{1}{4\pi} h^2
	+ \frac{\pi^3}{40} t^4 +\frac{1}{64\pi} h^4
	+\frac{\pi}{16} t^2 h^2 ;\\
w_{2b} = \;&
	-\frac{4}{3\pi^2} -\frac{5}{18} t^2 -\frac{11}{24\pi^2} h^2
	- \frac{17\pi^2}{1440}t^4 +\frac{47}{512\pi^2} h^4
	+ \frac{35}{192} t^2 h^2.
}
\label{omegaAnalyticN2}
\end{subequations}

\subsection{The Two-Bubble Diagram $\Omega_{2a}$}

\label{twoBubbleSection}

The term $\Omega_{2a}$ (Figure \ref{omega2a}) is unique among all
vacuum diagrams: depending on how the diagram is arranged, the
scattering process involved can be seen as taking place in any one
of the three channels: scattering of a particle-hole pair of the same
spin, scattering of a particle-hole pair of different spins, and
scattering of a particle-particle pair.  The first two corresponds to
the ``ring'' and ``ladder'' classifications, respectively.  The
particle-particle Cooper channel is linearly divergent in the UV, and
we implicitly subtract off the diverging part.  In the remainder of
this section, we will consider the diagram exclusively in the ring
configuration.  Setting $\N = 2$, the Feynman diagram in Figure \ref{ring2} yields:
\beq
\frac{\Omega_{2a}}{V} =
	- \left( \frac{4 \pi a}{m}\right)^{\! 2} \! \!
	\int \prod_{i = 1}^{3} \frac{\intd^3 \bm{k}_i}{(2\pi)^3}
		\frac{
			n_{\uparrow}(k_1) n_{\downarrow}(k_2)
			\left[ n_{\uparrow}(k_3) + n_{\downarrow}(k_4)\right]
			}
			{
			\frac{1}{2m}\left[ k_1^2 + k_2^2 -k_3^2 -k_4^2 \right]
			},
\label{twoBubble}
\eeq
\end{widetext}
where $\bm{k}_1 + \bm{k}_2 - \bm{k}_3 - \bm{k}_4 = 0$ by momentum
conservation.

From here one identifies the quantity $w_{2a}$ as defined in
\eqref{dimlessN2}:
\beq
w_{2a}(t, h) =
	\frac{6 \pi^2 m}{k_{\mu}^7 a^2} \frac{\Omega_{2a}(T, H)}{V}.
\label{omega2aN2}
\eeq

In the prequel, this term was examined numerically at $H = 0$.
It has the form:
\beq
w_{2a}(t, 0) =
	 B + C_1 \, t^2 + D_1 \, t^4 \ln t + E_1 \, t^4 + \dots
\label{omega2aT}
\eeq

On the other hand, at zero temperature the integral \eqref{twoBubble}
can be done analytically \cite{Kanno1970}, yielding
\beq
w_{2a}(0, h) =
	B + C_2 \, h^2 + D_2 \, h^4 \ln \vert h \vert
	+ E_2 \, h^4 + \dots
\label{omega2aH}
\eeq

Next, we attempt the double expansion, first in $t$ and then in $h$.
The result is of the form:
\beq
w_{2a}(t, h) = 
	B + C_1 \, t^2 + C_2 \, h^2 + f_4(t, h) + (\text{sixth order}),
\label{omega2aFull}
\eeq
where the fourth order term $f_4$ is
\beq
\spl{
f_4(t, h) =& \; 
	 \frac{1}{2} D_1 \, t^4
		\left( \ln \vert h \vert +  \ln \Delta \right)
	+ D_2 \, h^4 \ln \vert h \vert \\
	& + F_1 \, t^4 + E_2 \, h^4 + F_3 \, t^2 h^2 \\
	& + t^4 \left[
		\xi\!\left( t, \Delta \right) +
		\sum_{i = 1}^{\infty}
			F_{4,i} \left( \frac{t}{h} \right)^{2i}
		\right].
}
\label{4thOrder}
\eeq
Here $\Delta$ is an infinitesimal infrared cutoff imposed on the
momentum transfer $\bm{q}_r = \bm{k}_1 - \bm{k}_3$ to regularize
the results.  (See the appendix \ref{sommerfeldAppen} for the
complication of a momentum cutoff.)

Apart from the infinitely many $F_{4, i}$, all coefficients appearing
in \eqref{omega2aT}, \eqref{omega2aH} and \eqref{4thOrder} are given
in table \ref{w2aCoefficients}.

\begin{table}
\begin{tabular}{c|c}
$B$ & $\displaystyle \frac{4 (11 - 2 \ln 2)}{35\pi^2}$ \\ 
\\ [-0.5em]
$C_1$ & $\displaystyle -\ln 2 + \frac{1}{2}$ \\ 
\\ [-0.5em]
$C_2$ & $\displaystyle -\frac{(1 + 2 \ln 2)}{8\pi^2}$ \\ 
\\ [-0.5em]
$D_1$ & $\displaystyle -\frac{\pi^2}{10}$ \\ 
\\ [-0.5em]
$D_2$ & $\displaystyle \frac{1}{32\pi^2}$ \\ 
\\ [-0.5em]
$E_1$ & $1.62$ \\ 
\\ [-0.5em]
$E_2$ & $\displaystyle \frac{29 - 102 \ln 2}{1536\pi^2}$ \\ 
\\ [-0.5em]
$F_1$ & $\displaystyle \frac{6597 - 704 \ln 2}{38400 \pi^2}$ \\ 
\\ [-0.5em]
$F_3$ & $\displaystyle \frac{19 - 6 \ln 2}{64}$ \\ 
\\ [-0.5em]
$\kappa$ & $1.70$ \\ 
\end{tabular} 
	\caption{\label{w2aCoefficients} The coefficients for of
	$w_{2a}$.  The value of $E_1$ comes from the numerical fit
	in the prequel \cite{Cheng2017}.  The value of $\kappa$ is
	determined by equation \eqref{kappa}.}
\end{table}

Instead of a non-analytic $t^4 \ln t$ term, we found
$t^4 \left(\ln h + \ln \Delta \right)$.  Going to higher orders
in $T$, we found increasingly singular terms with powers of $h$
and $\Delta$ in the denominator, combining to an overall fourth
order.  We are naturally unable to carry this calculation to
infinite order in $T$, but it is not difficult to infer, using
dimensional analysis and symmetry argument, that the $h$ part forms
an infinite series of $(t/h)^{2n}$.

In \eqref{4thOrder}, we denote the $\Delta$ counterpart of these
higher singular terms as $\xi(t, \Delta)$.  We made no attempt to
infer a general form of $\xi$.  But it is clear that the original 
integral \eqref{twoBubble} is finite.  One therefore concludes that
all infrared divergent terms must resum into a finite quantity; that
is,
\beq
\frac{1}{2} D_1 \ln \Delta + \xi(t, \Delta)
	= \frac{1}{2} D_1 \ln t + \kappa,
\label{IRresum}
\eeq
where $\kappa$ is a constant yet unknown.  It will be determined
later by matching with the numerical result \eqref{omega2aT}.

Note that when one sets $t$ to zero, $f_4(0, h)$ as given in 
\eqref{4thOrder} reduces to $D_2 h^4 \ln h + E_2 h^4$, in exact
agreement with \eqref{omega2aH}.

By imposing an upper cutoff in $\vert \bm{q}_r \vert$ that is
\emph{smaller} than the Fermi momentum, we also verified that the
$t^4 \ln h$ term has contribution only from
$2 \sqrt{2m \mu_{\downarrow}} < \vert q_r \vert
< 2 \sqrt{2m \mu_{\uparrow}}$.  That is, it comes from the region
where $\vert q_r \vert \approx 2 k_F$, confirming the earlier claim
in the literature \cite{Chubukov2006, Maslov2009}.

The absence of a logarithmic term with the $t^2 h^2$ prefactor in
\eqref{4thOrder} is notable.  The accepted wisdom \cite{Carneiro1977}
is such that the spin susceptibility \emph{does not} scale as
$T^2 \ln T$, which is in line with our result here.  Granted, in
the present form \eqref{4thOrder} is only appropriate when
$t/h \ll 1$, while spin susceptibility is defined near zero magnetic
field.  But the coefficients to the logarithmic terms are robust:
resummation of the series $(t/h)^{2i}$ cannot generate a separate
$t^2 h^2$ logarithmic term.  If it is absent for $t/h \ll 1$, it
must remain so for all values of the ratio.

\section{Resumming the Singular Terms: Fermi Liquid Picture}

The original loop integral \eqref{twoBubble} is finite when $h$ is set
to zero.  Yet the expression \eqref{4thOrder} is not even well-defined
in the same limit.

To obtain a well-defined expression for $f_4$ in this limit, in
principle one only needs to swap the order of $h$- and $t$-expansion.  
Unfortunately, we cannot analytically evaluate the resultant integrals.
Instead, we will identify the $2k_F$ non-analyticity of \eqref{twoBubble}
with the infrared non-analyticity of the ladder diagram (figure
\ref{ladder2}), and evaluate the latter exactly
within the FL picture.

\subsection{The equivalence between Ladder and $2k_F$ Non-analyticity}

Historically, the study of non-analyticity of FL was framed in terms
of the quasiparticle self-energy.  Amit, Kane and Wagner
\cite{Amit1968} were the first to observe that interaction with a
particle-hole pair at either zero or $2k_F$ momentum results in the
leading logarithmic correction to the self-energy.  In a lengthy
paper, Chubokov and Maslov \cite{Chubukov2003} established that the
non-analytic contribution from the scattering of a particle-hole pair
at $2k_F$ is exactly equivalent to that of a \emph{particle-particle}
pair at zero momentum.  Noting that the proper self-energy is obtained
by differentiating vacuum 2PI Feynman diagrams, the above result
essentially constitutes a proof that, for our $\Omega_{2a}$ term
(see figure \ref{omega2a}), the $\vert q_l \vert = 0$ non-analyticity
is equivalent to that of $\vert q_r \vert \rightarrow 2k_\mu$.  
Nevertheless we will offer a standalone argument here, applied 
specifically to $\Omega_{2a}$.

Non-analyticity of $\Omega_{2a}$ must come from where the integrand
in \eqref{twoBubble} is singular at $T = H = 0$.  One notes that the
denominator of the integrand is proportional to
$(\bm{q}_l \cdot \bm{q}_r)$.  It may then appear that there are three
separate cases: $\bm{q}_l = 0$, $\bm{q}_r = 0$, and
$\bm{q}_l \perp \bm{q}_r$.  However, one can find a suitable change
of variables such that the integration measure transforms as
\beq
\intd^3 \bm{k}_1 \, \intd^3 \bm{k}_2 \, \intd^3 \bm{k}_3 
	\rightarrow 
	\intd^3 \bm{q}_r \, \intd^3 \bm{q}_l \, \intd^3 \! \bm{X},
\eeq
where $\bm{X}$ is some way to represent the remaining three degrees
of freedom.  It is clear that the vanishing integration measure
renders the limits $\bm{q}_l \rightarrow 0$ and
$\bm{q}_r \rightarrow 0$ regular by themselves.  The only actual
singularity is where $\bm{q}_l$ and $\bm{q}_r$ are orthogonal to
each other.

We have shown at the end of section \ref{twoBubbleSection} that the
non-analytic terms come from either
$\vert \bm{q}_r \vert \rightarrow 0$ or
$\vert \bm{q}_r \vert \rightarrow 2 k_\mu$, and let us concentrate
on the $2 k_\mu$ condition.  The region of the momentum integration
that contributes to the non-analyticity must then satisfy both the
$2 k_\mu$ and the orthogonality conditions.

\begin{figure}
\includegraphics[width = 0.4\textwidth]{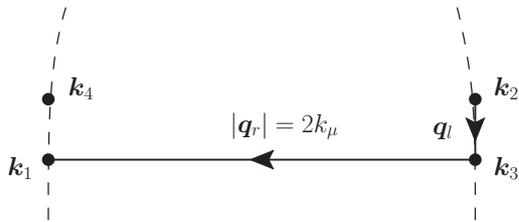}
	\caption{\label{nonAnalyticOriginFigure} The configuration
	around which the $2 k_{\mu}$ non-analyticity of $\Omega_{2a}$
	comes from.  The dashed lines represent the Fermi sphere.  The
	requirements are $\vert\bm{q}_r\vert \rightarrow 2 k_{\mu}$,
	$\bm{q}_l \perp \bm{q}_r$, and that $\bm{k}_1, \ldots \bm{k}_3$
	lie within the Fermi sphere.  Consequently $\bm{q}_l$ must
	vanish, and all four momenta $\bm{k}_1, \ldots \bm{k}_4$ must
	be exactly on the Fermi surface.}
\end{figure}

Let us now analyze the part of \eqref{twoBubble} that is proportional
to $n_{\uparrow}(k_1) \, n_{\downarrow}(k_2) \, n_{\uparrow}(k_3)$.
At zero temperature and magnetic field, these Fermi functions all
become the step function, restricting $\bm{k}_1$, $\bm{k}_2$ and
$\bm{k}_3$ to within the Fermi sphere.  The $2 k_{\mu}$ condition
forces $\bm{k}_1$ and $\bm{k}_3$ to sit exactly on the Fermi surface,
and be polar opposite to one another.  Note that by momentum
conservation $\bm{q}_l = \bm{k}_3 - \bm{k}_2$.  The orthogonality
condition then forces $\bm{q}_l$ to vanish.  See Fig 
\ref{nonAnalyticOriginFigure} for illustration.

For the other part of \eqref{twoBubble} where $n_{\uparrow}(k_3)$
is replaced by $n_{\downarrow}(k_4)$, one can identify
$\bm{q}_r = \bm{k}_2 - \bm{k}_4$ and
$\bm{q}_l = \bm{k}_1 - \bm{k}_4$, and the same argument goes
through.

One can reverse the argument to show that the
$\bm{q}_l \rightarrow 2 k_{\mu}$ non-analyticity is always paired
with the limit $\bm{q}_r \rightarrow 0$.  It is thus concluded that
non-analyticity of $\Omega_{2a}$ comes from the limiting regions of
momentum integration where one of $\bm{q}_r$ and $\bm{q}_l$ vanishes,
and the other approaches $2 k_{\mu}$.

This can be thought of as a duality between the ring and ladder
diagrams (see Fig. \ref{omega2a}): the $2k_\mu$ non-analyticity in
the ring form is exactly the zero-momentum non-analyticity in the
ladder form, and vice versa.

Finally we note that, in either limit, $\bm{k}_4$ is forced to sit on
the Fermi surface too by momentum conservation.  Therefore, to capture 
the non-analyticity of $\Omega_{2a}$, it suffices to consider the
small-momentum limit of both ring and ladder diagrams in Fig
\ref{omega2a}, with all four fermion legs restricted to be near the
Fermi surface.

\subsection{Ladder Diagram and Fermi Liquid Approximation}

Consider the ladder Feynman diagram (Fig \ref{ladder2}), which gives
the term $\Omega_{2a}^{ab}$ before summing over spins.  The Feynman 
diagram can be understood as the trace of the square of particle-hole 
Green's function $\Pi_{ab}(i\nu, q)$, where $\nu$ is the Bosonic 
Matsubara frequency.  This reduces to equation \eqref{twoBubble} if
the full non-interacting form of $\Pi_{\updown}$ is used.

However, our present goal is to compute all non-analytic terms
\emph{not} coming from $\vert q_r \vert \rightarrow 0$, and the
preceding discussion made clear that one only needs to look at a
limit: $\vert \bm{q}_l \vert \rightarrow 0$, with
$\bm{k}_1, \ldots \bm{k}_4$ all near the Fermi surface.  One can
then employ the asymptotic form of $\Pi_{\updown}$ for small $q$,
denoted $\pi_{\updown}$:
\beq
\pi_{\updown}(\ve, q) =
	\frac{k_{\mu}^2}{(2\pi)^2 v_{\mu}}
	\left\lbrace
		2 + \frac{\ve}{v_{\mu} \, q} \ln \left[
			\frac{\left( \frac{\ve + H}{v_{\mu} \, q} \right) -  1}
				{\left( \frac{\ve + H}{v_{\mu} \, q} \right) + 1}
		\right]
	\right\rbrace .
\label{pair}
\eeq

Corrections to this approximated form come in as positive powers of
$(H/v_{\mu} \, k_{\mu})^2$, $(T/v_{\mu} \, k_{\mu})^2$, or
$(q/k_{\mu})^2$.  Since $k_{\mu}$ is to be viewed as an ultraviolet
scale of the FL effective theory, these corrections are to be regarded
as vanishingly small in the present approximation.  It is also not
difficult to see that these beyond-FL corrections only yield overall
sixth order terms and higher if they are included in the following 
calculation.

We define a modified $\tilde{w}_{2a}$ based on $w_{2a}$, with the
full particle-hole bubble $\Pi_{\updown}$ replaced by
$\pi_{\updown}$:

\begin{widetext}

\beq
\tilde{w}_{2a}(t, h) =
	- \left(\frac{24 \pi^4}{k_{\mu}^7 m} \right) \, 
	T \sum_{\nu} \int^{\Lambda} \!\!
		\frac{\intd^3 q}{(2 \pi)^3} \; 
			\left( \pi_{\updown}(i \nu, q)^2
			+ \pi_{\downarrow \! \uparrow}(i \nu, q)^2
		\right) .
\label{wFL}
\eeq
This quantity contains the contribution to $w_{2a}$ coming from the
$q_l \rightarrow 0$ region of the momentum integral.  Due to the
small-$q$ approximation, a large-momentum cutoff $\Lambda$ is
necessary to render the expression finite.  We assume the hierarchy
of scales: $v_{\mu} k_{\mu} \gg v_{\mu} \Lambda \gg T, H$.

With a convergence factor $e^{i \omega o^{+}}$ appended, the Matsubara
frequency sum in \eqref{wFL} can be carried out using standard contour 
integration tricks.  The sum is transformed into an integral over
energy $\ve$, weighted by the usual Bose function
$n_B(\ve) = (e^{\beta \ve} - 1)^{-1}$, around the branch cut of
$\pi_{ab}(\ve, q)^2$ on the real axis.  The energy integral only
picks up the discontinuity of the integrand across the branch cut,
which is precisely the imaginary part.  The integration in $\bm{q}$
can also be carried out.

The end result contains a number of analytic terms: $T^4$, $H^4$,
$T^2 H^2$, $\Lambda^2 T^2$, $\Lambda^2 H^2$, and $\Lambda^4$.  The
cutoff $\Lambda$ appears here because the omitted large-$q$ processes
can nonetheless have small energy, and contribute equally well at
order $T^2$ and $H^2$.  The cutoff dependence signals the
incompleteness of our treatment.  This nevertheless poses no problem,
as we only want to capture the non-analytic terms with this approach.
The non-analytic part of
\eqref{wFL} reads:
\beq
\spl{
\tilde{w}_{2a}(t , h) = 
	-\frac{3 m^4}{\pi^2 k_{\mu}^8}  \left \lbrace \,
		2 \! \int_{0}^{\infty} \!\! \intd \ve \; n_B(\ve) \, \ve^2
			(\ve + H) \ln
				\left\vert \frac{\ve+H}{v_{\mu} \Lambda} \right\vert 
	+ \int_{0}^{v_F \Lambda - H} \!\! \intd \ve \; \ve^2
		(\ve + H) \ln \left\vert
			\frac{\ve + H}{v_{\mu} \Lambda}
		\right \vert
	\right \rbrace \\
+ (H \rightarrow -H) + (\text{analytic}).
}
\eeq

The second term above can be integrated to yield
$h^4 \ln \vert H/ v_{\mu} \Lambda \vert$ plus analytic terms.  In
addition, one can split
$\ln \vert(\ve + H)/v_{\mu} \Lambda \vert =
\ln \vert\ve/v_{\mu} \Lambda\vert + \ln \vert(\ve + H)/ \ve \vert$
in the integrand and notes that
\beq
\frac{1}{2} \, D_1 t^4 \ln \left( \frac{T}{v_{\mu} \Lambda} \right) =
- \frac{6 m^4}{\pi^2 k_{\mu}^8} 
	 \int_{0}^{\infty} \!\! \intd \ve \; n_B(\ve) \, \ve^3
		\ln \left\vert \frac{\ve}{v_{\mu} \Lambda} \right\vert + (\text{analytic}).
\eeq
This relation can be used to write $\tilde{w}_{2a}$ in the following form:
\beq
\tilde{w}_{2a}(t , h) = \chi(t, h) + \frac{1}{2} D_1 \, t^4 \ln t
	+ D_2 \, h^4 \ln h + (\text{analytic}),
\label{wFLNonAnalytic}
\eeq
where the \emph{crossover function} $\chi(t, h)$ is defined as
\beq
\chi(t, h) =
	-\frac{6 m^4}{\pi^2 k_{\mu}^8}
		\int_{0}^{\infty} \!\! \intd \ve \; n_B(\ve) \, \ve^2 \left[
			(\ve + H) \ln
				\left\vert \frac{\ve+H}{\ve} \right\vert
			+ (\ve - H) \ln
				\left\vert \frac{\ve-H}{\ve} \right\vert
		\right].
\label{crossoverChi}
\eeq

\end{widetext}

The expression \eqref{wFLNonAnalytic} and \eqref{crossoverChi}
capture \emph{all} the ladder-type non-analytic terms, which we
were unable to obtain to infinite order using the Sommerfeld
expansion approach in \eqref{4thOrder}.  Indeed, both $t^4 \ln t$
and $h^4 \ln \vert h \vert$ are recovered with correct coefficients.

In order to make comparison with \eqref{omega2aFull} and 
\eqref{4thOrder}, one needs to evaluate $\chi(t, h)$ in the limit
where the ratio $\alpha = t/h \ll 1$.  This is accomplished by
expanding out the logarithm in \eqref{crossoverChi}, assuming $\ve/H$
is always small.  The expansion is justified because the Bose function
$n_B(\ve)$ allows only contribution from the range $\ve < T$.  The
result is:
\beq
\spl{
\chi(t, h) =& \; \frac{1}{2} D_1 \, t^4
	\ln\left(\frac{h}{t}\right)\\
	&+ t^4 \left[
		\frac{\pi^4(5 - 6 \gamma_E) + 540 \, \zeta'(4)}{120\pi^2}
		+ \order{\alpha^2}
	\right].
}
\label{crossoverChiLargeH}
\eeq

One immediately notes that the $\ln\vert h/t \vert$ term correctly
converts the $\ln t$ in \eqref{wFLNonAnalytic} into
$\ln \vert h \vert$ in the limit $\alpha \rightarrow 0$ where $h$
dominates over $t$.

Finally one can compare \eqref{wFLNonAnalytic} and 
\eqref{crossoverChiLargeH} with \eqref{omega2aFull} and
\eqref{4thOrder}, and identify the $\order{\alpha^2}$ terms in
\eqref{crossoverChiLargeH} with the infinite series in
\eqref{4thOrder}.  This yields:
\beq
\spl{
\sum_{i=1}^{\infty} F_{4, i} \alpha^{2i} =&\;
	\frac{\chi(t, h)}{t^4} + \frac{1}{2} D_1 
		\ln\left(\alpha\right)\\
	&- \left(
		\frac{\pi^4(5 - 6 \gamma_E) + 540 \, \zeta'(4)}{120\pi^2} \right).
}
\label{resummation}
\eeq

As an extra check, we computed the coefficient $F_{4,1}$ both using
Sommerfeld expansion and from the crossover function $\chi(t, h)$.
Both approaches give identical answer $F_{4, 1} = \pi^4/63$.

\subsection{$\Omega_{2a}$ Near The $T$-axis }

The infinite series in \eqref{4thOrder} has been resummed using 
\eqref{resummation}.  It is natural to ask if one can now find a
well-defined expression for $w(t, h)$ when the ratio $\alpha$ is
large.

As it turns out, it is quite tricky to expand $\chi(t, h)$ around a
small $1/\alpha$.  We relegate the detail to the appendix, and note
the result here:
\beq
\spl{
\chi(t, h) =& \, 
	D_2 \, h^4 \ln \! \left(\frac{t}{h}\right)
		- \frac{1}{16} t^2 h^2 \\
	&+ h^4 \left[ \frac{12(\ln 2\pi - \gamma_E) + 7}{384 \pi^2}
		  + \order{\!\frac{1}{\alpha^2}\!} \right]  .
}
\label{crossoverChiLargeT}
\eeq

Equations \eqref{resummation} and \eqref{crossoverChiLargeT} can be
substituted into \eqref{4thOrder} to give an expression of
$f_4(t, h)$ well-defined in the limit of large $\alpha$.  In
particular, we are finally in a position to determine the constant
$\kappa$ appearing in \eqref{IRresum}.  Matching the coefficient of
$t^4$ terms, one obtains:
\beq
E_1 = F_1 - \left(
		\frac{\pi^4(5 - 6 \gamma_E) + 540 \, \zeta'(4)}{120\pi^2}
		\right)
		+ \kappa,
\label{kappa}
\eeq
which yields $\kappa = 1.51$.

With this final piece of puzzle found, one has the complete
non-analytic equation of state of a dilute Fermi gas up to overall
fourth order in $t$ and $h$.  The fourth order term $f_4(t, h)$ as 
defined in \eqref{omega2aFull} is non-analytic, and its series
expansion takes different forms depending on the size of
$\alpha = t/h$.

When $\alpha \lesssim 1$, one has
\beq
\spl{
f_4(t, h) =& \; \frac{D_1}{2} \, t^4 (\ln h + \ln t) 
	+ D_2 \, h^4 \ln h \\
	&+ (F_1 + \kappa) t^4 + E_2 h^4 + F_3 t^2 h^2 \\
	&+ t^4 \sum_{i=1}^{\infty} \left( F_{4,i} \, \alpha^{2i} \right).
}
\label{f4LargeH}
\eeq
The coefficients $F_{4,i}$ for arbitrary $i$ can be computed using 
\eqref{resummation}.  Or even better, one can just numerically
evaluate $\chi(t, h)$ to resum the series.

One can obtain the corresponding expansion for $\alpha \gtrsim 1$
using \eqref{crossoverChiLargeT}.  The result is
\beq
\spl{
f_4(t, h) =& \; D_1 \, t^4 \ln t + D_2 \, h^4 \ln t \\
	&+ E_1 t^4 + G_2 h^4 + G_3 t^2 h^2 \\
	&+ h^4 \sum_{i=1}^{\infty} \left( G_{4,i} \, \alpha^{-2i} \right),
}
\label{f4LargeT}
\eeq
with $G_2 = E_2 + \frac{12(\ln 2\pi - \gamma_E) + 7}{384 \pi^2}$ and
$G_3 = F_3 - \frac{1}{16}$.  The infinite series in \eqref{f4LargeT}
sums to:
\beq
\spl{
\sum_{i=1}^{\infty} G_{4,i} \, \alpha^{-2i}
	=&\; \frac{\chi(t, h)}{h^4}
		- D_2 \,\ln \! \left(\alpha\right) + \frac{1}{16} \alpha^2 \\
	& - \left[ \frac{12(\ln 2\pi - \gamma_E) + 7}{384 \pi^2} \right].
}
\label{resummationLargeT}
\eeq

\section{The Crossover Behavior}

\begin{figure}
\includegraphics[scale=0.5]{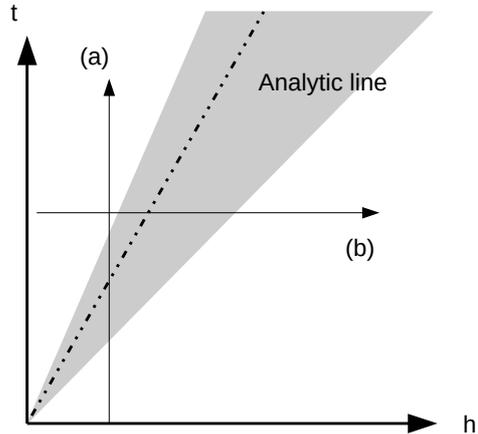}
	\caption{\label{phaseDiagram} This is a sketch of the
	crossover behavior of the ladder-type non-analyticity.  
	The shaded crossover region separates the two near-axis
	areas, where one scale dominates the other.  The analytic
	line in the crossover region is where the two sets of
	non-analyticity precisely cancel each other out.  The two
	paths (a) and (b) will be considered in subsequent
	discussion.}
\end{figure}

As discussed in section \ref{origin}, the ladder-type non-analyticity
(figure \ref{ladder2}) can be cut off in the infrared by either $T$
or $H$.  The competition between the two scales results in a
non-trivial crossover.  We sketch the behavior in figure 
\ref{phaseDiagram}.

Consider the expression \eqref{omega2aFull} for $w$.  The
non-analytic $f_4$ term admits two different expansions
\eqref{f4LargeH} and \eqref{f4LargeT}, good for the regions near the
$t$- and $h$-axis in figure \ref{phaseDiagram}, respectively.  When
the ratio $\alpha = t/h$ is neither larger nor small, the higher
order terms are important in both expansions, and this corresponds
to the shaded crossover region.  Fortunately, the series can be resum
exactly using \eqref{resummation} and \eqref{resummationLargeT}.

\subsection{Path (a): Raising $t$ with a fixed $h$}

\begin{figure}
\begin{subfigure}{0.45\textwidth}
	\includegraphics[width=1\textwidth]{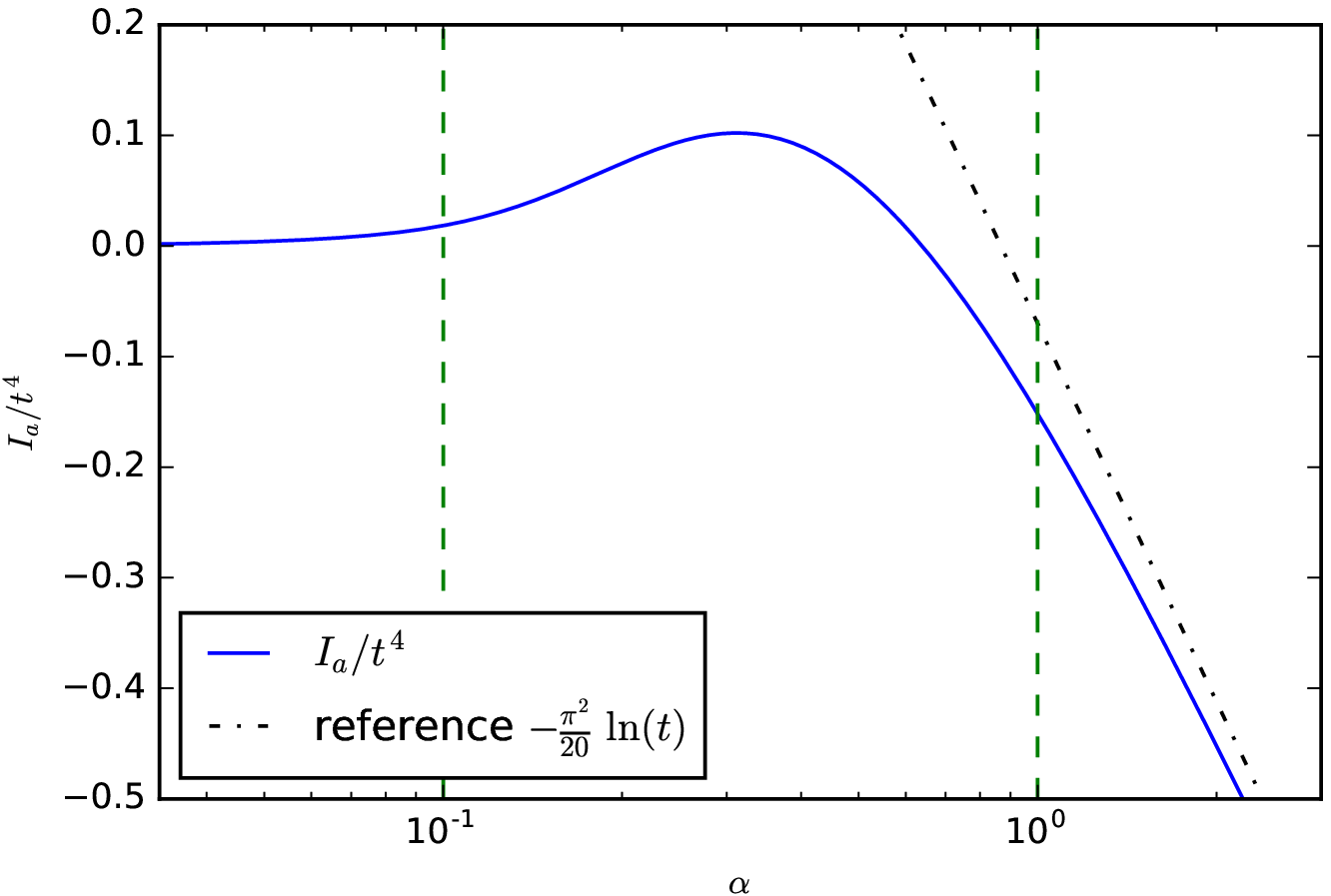}
	\caption{\label{crossoverPlotA}}
\end{subfigure}

\begin{subfigure}{0.45\textwidth}
	\includegraphics[width=1\textwidth]{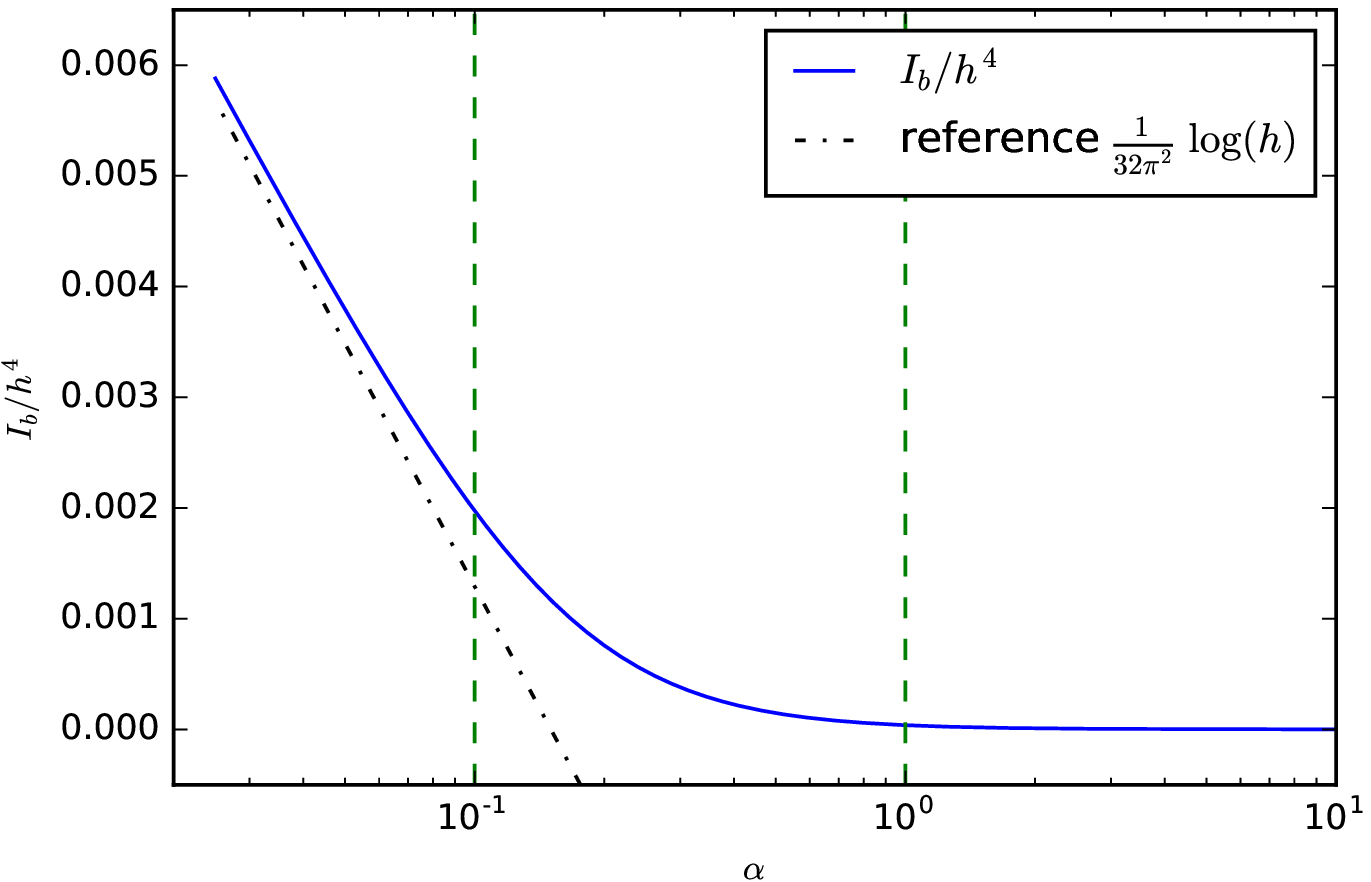}
	\caption{\label{crossoverPlotB}}
\end{subfigure}

	\caption{The plots of the ladder-type non-analytic parts
	(a) $I_a/t^4$ and (b) $I_b/h^4$ against $\ln \alpha$.
	(a) is up to a vertical offset of $\ln h$, while (b) is up
	to a vertical offset of $\ln t$.  The green dashed lines
	indicate the crossover regions on each plot, corresponding
	to the shaded area in FIG \ref{phaseDiagram}.  Each curve
	approaches zero on one side of the crossover region, indicating
	the asymptotic $t^4 \ln h$ and $h^4 \ln t$ behaviors of $I_a$
	and $I_b$, respectively.  On the opposite sides of the crossover
	region, $I_a$ and $I_b$ respectively become $t^4 \ln t$ and
	$h^4 \ln h$, as indicated by the asymptotically linear behavior.}
\end{figure}

This is perhaps the scenario most relevant to the experiment.  As
the effective magnetic field is controlled via the number densities
of individual spin components, moving along path (a) in figure 
\ref{phaseDiagram} corresponds to fixing the composition of the gas
and tuning the temperature.

In region near the $h$-axis, \eqref{f4LargeH} is the appropriate 
expression to use.  To identify the ladder-type non-analyticity, from
the thermodynamic potential one subtracts off all the analytic terms,
and half of the total $t^4 \ln t$ term associated with the ring-type
non-analyticity.  The result reads:
\beq
I_{a}(t, h) =  \frac{D_1}{2} \, t^4 \ln \vert h \vert
	+ t^4 \, \sum_{i=0}^{\infty} F_{4, i} \, \alpha^{2i}.
\eeq
We have taken the liberty to subtract off $D_2 h^4 \ln h$, which is
a constant along the path.  When $\alpha$ is small, $I_a$ approaches
$\frac{D_1}{2} t^4 \ln h$.  When $\alpha$ grows large, however, using
\eqref{resummation} and \eqref{crossoverChiLargeT}, one deduces
$I_a \sim  t^4 (\frac{D_1}{2} \ln t + \order{\alpha^{-2}})$.

For intermediate value of $\alpha$, the crossover can be followed by
numerically integrate $\chi(t, h)$.  We plot $(I_a/t^4)$ against
$\ln \alpha$ in figure \ref{crossoverPlotA}.  The crossover region can
be identified from the plot as $0.1 < \alpha < 1$, where the behavior
of the function $I_a$ substantially deviates from either asymptotic
forms.

The result of this section also answers a dangling question from the
prequel: the fate of the $t^4 \ln t$ behavior in an $\mathrm{SU}(\N)$
Fermi gas when the $\mathrm{SU}(\N)$ symmetry is broken by unequal
number densities of spin components.  The ring contribution to the
$t^4 \ln t$ term is wholly unaffected, while the ladder contribution
remains robust as long as the ratio $\alpha$ is of order unity or
bigger.

\subsection{Path (b): Raising $h$ with a fixed $t$}

For completeness, we consider this complementary scenario.  The 
appropriate expansion for $f_4$ is \eqref{f4LargeT} near the $t$-axis.  
After subtracting off the analytic terms and the $t^4 \ln t$ term for 
being constant along the path, one obtains the ladder-type
non-analyticity
\beq
I_{b}(t, h) =  D_2 \, h^4 \ln t
	+ h^4 \, \sum_{i=0}^{\infty} G_{4, i} \, \alpha^{-2i}.
\eeq

Using \eqref{crossoverChiLargeT} and \eqref{resummationLargeT}, one
sees that $I_b \sim h^4 ( D_2 \ln h + \order{\alpha^2})$ as $\alpha$
gets large.  A similar crossover plot of $(I_b/h^4)$ against
$\ln \alpha$ is presented in figure \ref{crossoverPlotB}.  The
crossover region $0.1 < \alpha < 1$ can be consistently identified
from the plot.

\subsection{The analytic line}

This is the most striking feature on figure \ref{phaseDiagram}, though 
perhaps also the hardest to access experimentally.  Consider the case 
where one takes both $t$ and $h$ to zero at fixed $\alpha$.  Let us 
define the ``Euclidean'' distance $l$ on the $(t, h)$-plane:
\beq
\spl{
l^2 \equiv& \; t^2 + h^2 \\
	=& \; h^2 \; (1 + \alpha^2) =  t^2 (1 + \frac{1}{\alpha^2}).
}
\eeq

It can be shown that $\chi$ as given in \eqref{crossoverChi} is of
the form $l^4 \tilde{\chi}(\alpha)$, where the function $\tilde{\chi}$
is independent of $l$.  Thus the ladder-type non-analyticity can be
cast into the following function:
\beq
\spl{
I(t, h) =&\;
	\chi(t, h) + \frac{D_1}{2} t^4 \ln t
	+ D_2 \, h^4 \ln \vert h \vert \\
	=& \; \left[\frac{D_1 \alpha^4 + D_2}{(1+\alpha^2)^2} \right]
	l^4 \ln l + \order{l^4}.
}
\eeq

The $l^4 \ln l$ term vanishes at the special ratio
\beq
\alpha_c =
	\pm \sqrt[4]{\left\vert \frac{D_2}{D_1} \right\vert} =
	\pm \sqrt[4]{\frac{8}{5}} \pi.
\label{alphaCritical}
\eeq

If one approaches $t = h = 0$ along this direction, the thermodynamic
potential appears as an entirely analytic function of the distance $l$
and, by extension, of $t$ or $h$.  The two sets of non-analytic
behaviors ``cancel'' each other out.

\subsection{Away from the dilute limit}

The preceding discussion assumes the diluteness condition.  As we have
argued that the logarithmic correction is well within the FL 
theory, the diluteness condition should not be essential for the
non-analytic crossover.  We will show that, by replacing the
interaction vertices in perturbation theory with the full
quasiparticle scattering amplitudes, one can write down the
non-analytic terms in a form that remains valid beyond the dilute
limit.

\begin{figure}
\includegraphics[width = 0.3\textwidth]{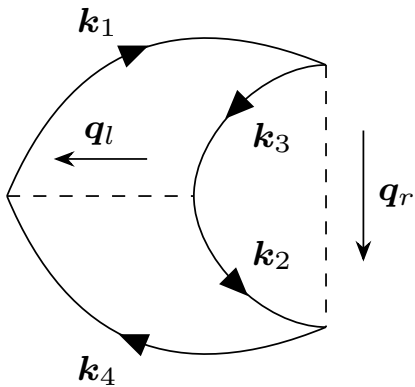}
	\caption{\label{crescent} The Feynman diagram corresponding to
	the new $\Omega_{2c}$ term.  It is arranged in the unconventional
	crescent shape to better expose the particle-hole pair structure.}
\end{figure}

In this section we instead consider a microscopic interaction
potential that couples through particle density (or, conventionally
for an electronic system, charge):
\begin{widetext}
\beq
H_{I} = \sum_{a, b} \sum_{\bm{p}, \bm{k}, \bm{q}}
			\frac{U(\vert q\vert)}{2}
			\psi^{\dag}_{a}\left(\bm{p}-\frac{\bm{q}}{2}\right)
			\psi^{\dag}_{b}\left(\bm{k}+\frac{\bm{q}}{2}\right)
			\psi_{b}\left(\bm{k}-\frac{\bm{q}}{2}\right)
			\psi_{a}\left(\bm{p}+\frac{\bm{q}}{2}\right).
\eeq
This alternative model directly allows for same-spin scattering.
Thus all scattering channels in the full FL phenomenology
are already present at first order in perturbation theory, and one
may directly extrapolate from there.  Because of the added possibility
of same-spin scattering, $\Omega_{2a}$ gains extra contribution, and
another vacuum diagram $\Omega_{2c}$ contributes to the non-analyticity 
at second order, as shown in Fig \ref{crescent}.

Assuming that $U(\vert q \vert)$ is analytic and positive everywhere,
the non-analyticity of $\Omega_{2a}$ still comes from the same
zero/$2 k_F$-momentum-transfer limits, and likewise for the new
$\Omega_{2c}$.  One is then allowed to make the approximation
$U(\vert q_{l,r} \vert) \approx U(0)$ or $U(2 k_{\mu})$ where
appropriate, and write
\beq
\frac{\Omega_{2a} + \Omega_{2c}}{V} =
	\left(\frac{k_{\mu}^7 m}{96 \pi^4}\right)
	\left \lbrace \left [
		2 \, U(0)^2 - 2 \, U(0)U(2 k_{\mu}) + U(2 k_{\mu})^2
		\right]
	\tilde{w}_{2a}(t, 0)
	+ U(2 k_{\mu})^2 \tilde{w}_{2a}(t, h)
	\right \rbrace
+ (\text{analytic}).
\label{stronglyInteractingNonAnalyticityBare}
\eeq
The non-analytic part of $\tilde{w}_{2a}$ is to be identified from
equation \eqref{wFLNonAnalytic}.

The so-called fixed point vertices $\Gamma_s$ and $\Gamma_c$ were
introduced in \cite{Chubukov2006}.  They represent the exact
scattering amplitudes of two quasiparticles, where the subscript $s$
and $c$ stands for charge and spin channel, respectively.  Here we
are only interested in the limits where $\bm{k}_1$ and $\bm{k}_3$ are
either equal or on opposite sides of the Fermi surface.  To the
lowest order in perturbation theory, the limiting values for the
scattering amplitudes are
\beq
\spl{
\Gamma_c =& \;
	\frac{m k_\mu}{\pi^2} \left[
		U(0)-\frac{1}{2}\, U(2 k_\mu)
	\right] ;\\
\Gamma_s
	=& \; -\frac{m k_\mu}{\pi^2} \frac{1}{2} \, U(2 k_\mu).
}
\label{flVertices}
\eeq

One may now identify $U(0)$ and $U(2k_{\mu})$ in 
\eqref{stronglyInteractingNonAnalyticityBare} with appropriate 
combinations of $\Gamma_c$ and $\Gamma_s$.  Higher order terms in
the perturbation theory will only serve to renormalize the values
of $k_{\mu}$, $m$, $\Gamma_c$ and $\Gamma_s$.  One then obtains an 
expression that remains valid outside the dilute regime:
\beq
(\text{non-analyticity}) =
	\frac{k_F^5}{96 m^{*}} \left \lbrace
		\left(
		\Gamma_s^2 +\Gamma_c^2 \right) \frac{D_1}{2} t^4 \ln t
		+ 2\Gamma_s^2 \left( \frac{D_1}{2}t^4 \ln t+
			D_2 h^4 \ln h + \chi(t, h)
		\right)
	\right \rbrace.
\label{stronglyInteractingNonAnalyticity}
\eeq
\end{widetext}

As advertised, \eqref{stronglyInteractingNonAnalyticity} only depends
on parameters in FL theory, and should be valid where the FL picture
holds.  One notes that the crossover is controlled solely by the
spin channel, as is intuitively expected.

We mention in passing that, under the zero/$2k_F$ ``duality'' that
map ladder and ring diagrams into each other, the new crescent diagram
$\Omega_{2c}$ is self-dual.

It should be noted that, beyond the dilute regime, one loses the
ability to calculate the analytic terms at fourth order, which we
have shown to mix with the non-analytic terms during the crossover.
(See equations \eqref{resummation} and \eqref{resummationLargeT}.)  
Therefore one no longer has a consistent approximation to the equation
of state.

Furthermore, while equation \eqref{stronglyInteractingNonAnalyticity} 
certainly remains valid, it captures only the process with two 
\emph{dynamical} particle-hole pairs.  As was shown in reference 
\cite{Chubukov2006}, a three-pair process also contributes to the
non-analyticity.  We were justified to ignore the term in the dilute
regime, but we acknowledge that the three-pair term is significant in 
general when the diluteness condition does not hold.

\section{Generalization to SU($\N$) }

\label{largeN}

When $\N > 2$, the additional complexity leads to a rich and exotic
phase diagram.  However, the meanfield treatment in the existing 
literature \cite{Cazalilla2009, Cazalilla2014} by construction yields
an analytic expression for the free energy.  The non-analytic effect 
discussed in the preceding section will qualitatively affect the phase
transition \cite{Belitz1999, Belitz2005, Maslov2009}.

In this section we give the equation of state including the
non-analytic effect, in the non-magnetic phase, generalized to
$\N > 2$.  We also propose an experimental scenario that offers the
advertised large-$\N$ enhancement of the non-analytic term.

\subsection{Equation of State}

\begin{table*}
\begin{tabular}{c|cccc}
\hline 
$i$ & $x=0$ & $1$ & $2a$ & $2b$ \\ 
\hline
\\ [-0.5em]
0 &
$\displaystyle -\frac{2}{5}$ &
$\displaystyle \frac{2}{3\pi}$ &
$\displaystyle \frac{4}{35\pi^2}(11 - 2 \ln 2)$ &
$\displaystyle -\frac{4}{3\pi^2}$ \\ 
\\ [-0.5em]
1 &
$\displaystyle -\frac{\pi^2}{4}$ &
$\displaystyle \frac{\pi}{6}$ &
$\displaystyle \frac{1}{2}(1-2 \ln 2)$ &
$\displaystyle -\frac{5}{18}$ \\ 
\\ [-0.5em]
2 &
$\displaystyle -\frac{3}{4\N}$ &
$\displaystyle \frac{1}{2\pi}\frac{\N-4}{\N(\N-1)}$ &
$\displaystyle \frac{1}{2\pi^2}
	\frac{(5\N-11)+(-\frac{3}{2}\N+1)\ln 2}{\N (\N-1)}$ &
$\displaystyle -\frac{1}{6\pi^2}
	\frac{5\N^2 -22\N + 35}{\N(\N-1)^2}$ \\ 
\\ [-0.5em]
3 &
$\displaystyle -\frac{1}{8\N}$ &
$\displaystyle -\frac{1}{12\pi}\frac{\N+8}{\N(\N-1)}$ &
$\displaystyle \frac{1}{2\pi^2}
	\frac{(\N-\frac{11}{2})+(-\frac{7}{8}\N +\frac{3}{2})\ln 2}
	{\N(\N-1)}$ &
$\displaystyle \frac{1}{12\pi^2}\frac{\N^2 -2\N -35}{\N(\N-1)^2}$ \\
\\ [-0.5em]
4 &
$\displaystyle \frac{7\pi^4}{960}$ &
$\displaystyle \frac{\pi^3}{40}$  &
$0$&
$\displaystyle -\frac{17\pi^2}{1440}$ \\ 
\\ [-0.5em]
5 &
$\displaystyle \frac{1}{64\N}$ &
$\displaystyle \frac{1}{32\pi}\frac{1}{\N-1}$ &
$\displaystyle \frac{1}{256\pi^2}
	\frac{(5\N-88)+(-13\N +8)\ln2}{\N(\N-1)} $
& $\displaystyle -\frac{1}{96\pi^2}
	\frac{\N^2 -6\N +35}{\N(\N-1)^2}$ \\ 
\\ [-0.5em]
6 &
$0$ &
$\displaystyle \frac{3}{32\pi}\frac{1}{\N(\N-1)}$ &
$\displaystyle \frac{1}{256\pi^2} \frac{39+9\ln 2}{\N(\N-1)} $ &
$\displaystyle -\frac{1}{16\pi^2}\frac{\N-16)}{\N(\N-1)^2}$ \\ 
\\ [-0.5em]
7 &
$\displaystyle -\frac{\pi^2}{32}\frac{1}{\N}$ &
$\displaystyle \frac{\pi}{8} \frac{1}{\N-1}$ &
$\displaystyle \frac{3}{32}\frac{(\N-2)(1-\ln 2)}{\N(\N-1)} $ &
$\displaystyle -\frac{1}{48}\frac{3\N^2-26\N+5}{\N(\N-1)^2} $ 
\end{tabular}
	\caption{\label{coefficientsTable} Numerical values of
	coefficients $a_{i}^{(x)}$ appearing in equations
	 \eqref{omegaAnalytic}} and \eqref{omega2aSpinSum}.
\end{table*}

It will be convenient to consider the dimensionless magnetic fields
$h_a = H_a/\mu_0$.  The thermodynamic potential must be
$\mathrm{SU}(\N)$-symmetric overall.  The analytic part can be
conveniently expressed in terms of these $\mathrm{SU}(\N)$ invariants:
\beq
S_n \equiv \sum_{a=1}^{\N} ( h_a )^n; \quad n = 2, \dots, \N.
\label{Sn}
\eeq
Any $n \geq \N$ term is a linear combinations of $S_2, \dots, S_{\N}$.
The first term $S_1$ vanishes by the traceless condition.  These 
quantities serve as ``monomials'' in a generalized power series 
expansion.

Generally speaking, $\mathbf{H} \rightarrow -\mathbf{H}$ is no longer
a symmetry of the model.  Thus, starting from $S_3$, odd terms are 
allowed in the expansion of the thermodynamic potential.  In the 
treatment of \cite{Cazalilla2009, Cazalilla2014}, the same physics 
manifests itself as odd powers of magnetization in the Ginzburg-Landau expansion of free energy.

Up to second order in perturbation theory, the generic expression for 
thermodynamic potential \eqref{dimless} is valid for arbitrary $\N$.
For $\Omega_0$, $\Omega_1$ and $\Omega_{2b}$, the spin sum can still
be carried out straightforwardly, and the results can be expressed in 
terms of $S_n$.  In analogy of \eqref{dimlessN2}, we define:
\begin{subequations}
\ali{
\omega_{0} =& \;
	\frac{1}{\N}\sum_{a = 1}^{\N} \omega_{0}^{(a)} ;\\
\omega_{1} =& \;
	\frac{1}{\N(\N-1)}\sum_{a = 1}^{\N} \sum_{b \neq a}
		\omega_{1}^{(ab)} ;\\
\omega_{2a} =& \;
	\frac{1}{\N(\N-1)}\sum_{a = 1}^{\N} \sum_{b \neq a}
	\omega_{2a}^{(ab)} ;\\
\omega_{2b} =& \;
	\frac{1}{\N(\N-1)^2}
	\sum_{a = 1}^{\N} \sum_{b \neq a} \sum_{c \neq a}
	\omega_{2b}^{(abc)}.
}
\label{omegaSymmetrized}
\end{subequations}

Let $x$ denotes $0$, $1$ or $2b$.  Up to the fourth overall order
in $t$ and $h$, these quantities have the general form
\beq
\spl{
\omega_{x} =& \;
a_0^{(x)} + a_1^{(x)} t^2 + a_2^{(x)} S_2 + a_3^{(x)} S_3
	+ a_4^{(x)} t^4 \\
&+ a_5^{(x)} S_4 + a_6^{(x)} (S_2)^2
	+ a_7^{(x)} t^2 S_2 + \dots
}
\label{omegaAnalytic}
\eeq
The coefficients $\lbrace a_i^{(x)} \rbrace$ are summarized in table 
\ref{coefficientsTable}.

For $\omega_{2a}$, it is easier to first withhold the spin sum, and
consider instead $\omega_{2a}^{(ab)}$ with definite spins $a$ and $b$.  
To this end, one defines the ``centered'' chemical potential
\beq
\mu_{ab} = \mu_0 + \frac{1}{2} \left( H_a + H_b \right),
\eeq
as well as related quantities $k_{ab}$, $v_{ab}$ and $t_{ab}$, where
one replaces all occurrences $\mu_0$ with $\mu_{ab}$ in the original 
definitions.  Also one defines $h_{ab} = (H_a - H_b)/\mu_{ab}$.

Since the dimensionless $\omega_{2a}^{(ab)}$ is symmetric under the 
exchange of $a$ and $b$ (obvious from the Feynman diagrams in Fig 
\ref{omega2a}), one may rewrite it with explicit spin symmetrization:
\beq
\omega_{2a}^{(ab)} =
	\frac{12 \pi^2 m}{k_{\mu}^7 a^2}
	\frac{\left(\Omega_{2a}^{(ab)} + \Omega_{2a}^{(ba)}\right)}{2V}.
\label{omega2aABSymmetrized}
\eeq

This nearly coincides with the $\N=2$ spin-symmetrized $w_{2a}$.  
However, to adopt the $\N = 2$ result, one must replace $\mu_0$ with
$\mu_{ab}$, with all the scales and dimensionless parameters modified 
accordingly.  The upshot is
\beq
\omega_{2a}^{(ab)}(t, h) =
	\left(\frac{k_{ab}}{k_{\mu}}\right)^{\!-7}
	w_{2a}\left(t_{ab}, h_{ab} \right).
\eeq

One may now carry out the spin average over all pairs $a, b$ in
\eqref{omega2aABSymmetrized}.  While the non-analytic terms cannot
be expressed with the $\mathrm{SU}(\N)$ invariants $S_n$ in a simple
way, the analytic part can still be summed.

The non-analytic part is not affected by the shift from $k_{\mu}$ to $k_{ab}$ at leading order:
\beq
\left(\frac{k_{ab}}{k_{\mu}}\right)^{\! 7} \! f_4(t_{ab}, h_{ab})
	= f_4(t, h_a - h_b) + (\text{sixth order}).
\eeq

In analogy to \eqref{omegaAnalytic}, the spin-symmetric $\omega_{2a}$
can be written as
\beq
\spl{
\omega_{2a} =& \;
a_0^{(2a)} + a_1^{(2a)} t^2 + a_2^{(2a)} S_2 + a_3^{(2a)} S_3\\
&+ a_5^{(2a)} S_4 + a_6^{(2a)} (S_2)^2
	+ a_7^{(2a)} t^2 S_2 \\
&+ \frac{1}{\N(\N-1)} \sum_{a}\sum_{b \neq a} f_4(t, h_a - h_b)
+ \dots
}
\label{omega2aSpinSum}
\eeq
The coefficients are also given in Table \ref{coefficientsTable}.
Note that the $t^4$ term is identically zero in the above expansion.

\subsection{Experimental Scenario:
	$\N \rightarrow \frac{\N}{2} + \frac{\N}{2}$}

We consider the experimental setup that forbids transitions among
$\mathrm{SU}(\N)$ spin states.  But the $\mathrm{SU}(\N)$ symmetry
is still broken by the unequal densities of spin components, which
is equivalent to a non-zero generalized magnetic field in our model.

To observe the non-analytic crossover, a magnetic field is obviously 
needed.  Yet we hope for an enhancement of the non-analytic effect,
which receives ``extra copies'' of the same contribution due to the 
unbroken part of the symmetry.  The simplest scenario works best to 
fulfill the requirements: we shall consider $\N$ even, and the
$\mathrm{SU}(\N)$ being broken neatly into
$\mathrm{SU}(\frac{\N}{2}) \times \mathrm{SU}(\frac{\N}{2})$.  This
corresponds to the effective magnetic fields:
\beq
H_a =
\begin{cases}
	H/2,	&\quad a \leq \frac{\N}{2}; \\
	-H/2, &\quad a > \frac{\N}{2}.
\end{cases}
\label{scenarioH}
\eeq

This particular scenario closely resembles the spin-$\frac{1}{2}$
electron gas, and offers the largest enhancement of the non-analytic 
terms.

Let $h = H/\mu_0$, similar to the $\N = 2$ case.  In this particular
scenario, the $\mathrm{SU}(\N)$-invariants $S_n$ defined in \eqref{Sn}
becomes

\beq
S_n =
\begin{cases}
	\N \left(\frac{h}{2}\right)^n  & \quad \text{$n$ even}; \\
	0, & \quad \text{$n$ odd}.
\end{cases}
\label{SnExperiment}
\eeq
The odd power terms vanish identically due to the restored
$H \rightarrow -H$ symmetry.

One can substitute \eqref{SnExperiment} into \eqref{omegaAnalytic}
and \eqref{omega2aSpinSum} to recover the equation of state.  The
full expression is very long and we shall not print it here, but we
draw attention to the non-analytic part of $\Omega_{2a}$:
\beq
\spl{
\sum_{a}\sum_{b \neq a}& \; f_4(t, h_a - h_b) \\
=& \; \frac{\N(\N-2)}{2} f_4(t, 0) + \frac{\N^2}{2} f_4(t, h)
}
\eeq

Both terms are proportional to $\N^2$, compared with the linear
scaling of the non-interacting part.  This is the potential
large-$\N$ enhancement that we hope can make the experimental
detection of the non-analytic behaviors less difficult.

\subsection{Hartree-Fock Resummation}

The above argument for the large-$\N$ enhancement is flawed, however.
From table \ref{coefficientsTable} one can see that both $\omega_{1}$
and $\omega_{2b}$ scale as $\order{1}$ when $\N$ is large.  After
spin sum, $\Omega_{1}$ also scales as $\order{\N^2}$, while
$\Omega_{2b}$ is of order $\order{\N^3}$!

Since we have been advocating the large-$\N$ enhancement, one may
question if this doesn't actually make $\Omega_{2a}$ \emph{less}
visible in an experiment.  In fact, at each order of perturbation
theory, the diagrams that form parts of the Hatree-Fock approximation
are always proportional to the highest possible power of $\N$.  As an
alternative, we propose that the Hartree-Fock terms may be resummed
using a scheme inspired by the familiar Luttinger-Ward (LW)
functional \cite{Luttinger1960}.

The lowest-order skeleton diagram for the LW scheme coincides with
the diagram for $\Omega_{1}$ (figure \ref{omega1}).  If one chooses
to include only this diagram, the LW scheme produces only a
spin-dependent constant shift $\sigma_a$ on top of the chemical
potential $\mu_a$.  We shall denote the resultant (approximated) 
thermodynamic potential as $\Omega_{HF}$:
\beq
\spl{
\frac{\Omega_{HF}}{V} =
	\left(\frac{m T}{2 \pi}\right)^{\frac{3}{2}} \sum_{a} \bigg[
		T \, \Li{\frac{5}{2}} \!
			\left( -e^{\beta(\mu_0 + H_a - \sigma_a)} \right) \\
		+ \frac{1}{2} \sigma_a \, \Li{\frac{3}{2}} \!
			\left( -e^{\beta(\mu_0 + H_a - \sigma_a)} \right)
	\bigg],
}
\label{HFThermodynamicPotentialExact}
\eeq
where $\Li{s}$ is the polylogarithm.

The usual stationary condition for the Luttinger-Ward functional
yields a self-consistent condition of the energy shifts:
\beq
\sigma_a = - \frac{4\pi a}{m}
	\left(\frac{mT}{2 \pi}\right)^{\frac{3}{2}}
		\sum_{b \neq a} \Li{\frac{3}{2}} \! \left(
			-e^{\beta(\mu_0 + H_b - \sigma_b)}
		\right).
\label{HFSelfConsistent}
\eeq

The above approximation exactly resums the Hartree-Fock self-energy
to all orders in perturbation theory; thus the subscript ``HF''.  One
can add to $\Omega_{HF}$ any beyond-HF vacuum diagram to further
refine the approximation.  In particular, we wish to write
\beq
\Omega \approx \Omega_{HF} + \Omega_{2a} +\order{(k_{\mu}a)^3}.
\label{omegaPhysical}
\eeq
We keep the perturbative power counting for beyond-Hartree-Fock
corrections, despite that the resummation leading to $\Omega_{HF}$
is already non-perturbative.

Experimentally, one may already solve \eqref{HFSelfConsistent} using
the measured values of $T$, $\mu_0$ and $H_a$.  And then one can 
calculate $\Omega_{HF}$ from \eqref{HFThermodynamicPotentialExact},
and subtract it off the measured value of $\Omega$ to expose
$\Omega_{2a}$.  However, we propose a further approximation scheme
that simplifies the analysis.

First, one notes that $N_a$, the physical number density of spin $a$,
satisfies the following relation:
\beq
N_a = - \left(\frac{mT}{2 \pi}\right)^{\frac{3}{2}}
		\Li{\frac{3}{2}} \! \left(
			-e^{\beta(\mu_0 + H_b - \sigma_b)}
		\right)
		+ \order{(k_{\mu}a)^2}.
\label{physicalDensity}
\eeq

Comparing this equation with \eqref{HFSelfConsistent}, and fropping the correction terms on the right hand side of \eqref{physicalDensity}, one may make the approximations:
\beq
\label{HFshift}
\sigma_a \approx \frac{4\pi a}{m} \sum_{b \neq a} N_b.
\eeq
And then $\Omega_{HF}$ itself can be approximated as:
\beq
\label{HFthermodynamicPotential}
\frac{\Omega_{\text{HF}}}{V} \approx
	\sum_{a = 1}^{\N} \left[ \, 
		\left( \frac{mT}{2 \pi} \right)^{\!\frac{3}{2}}
		T \, \Li{\frac{5}{2}}\! \left(
			-e^{\beta(\mu_0 + H_a - \sigma_a)}
		\right) 
		- \frac{ \sigma_a N_a}{2}
	\right].
\eeq

This approximation does away with the transcendental equation
\eqref{HFSelfConsistent}.  Crucially, the error introduced to
$\Omega_{HF}$ is only of order $\order{(k_{\mu} a)^3}$.  Therefore
\eqref{omegaPhysical} remains valid, and can be used to identify
$\Omega_{2a}$ experimentally.

For the present scenario \eqref{scenarioH}, the number density of
each spin component must satisfy
\beq
N_a =
	\begin{cases}
		N_0 + \frac{\Delta N}{2}	& a \leq \frac{\N}{2}, \\
		N_0 - \frac{\Delta N}{2}	& \text{otherwise.}
	\end{cases}
\eeq

Then the self-energy shift \eqref{HFshift} becomes
\beq
\sigma_a \approx \begin{cases}
		\frac{4 \pi a}{m} \left[
			\left(\N - 1\right) N_0 - \frac{\Delta N}{2}
		\right] \equiv \sigma_{\uparrow}
		& a \leq \frac{\N}{2}, \\
		\frac{4 \pi a}{m} \left[
			\left(\N - 1\right) N_0 + \frac{\Delta N}{2}
		\right] \equiv \sigma_{\downarrow}
		& \text{otherwise.} 
	\end{cases}
\eeq

And $\Omega_{HF}$ is
\begin{widetext}
\beq
\frac{\Omega_{HF}}{V} \approx \frac{\N}{2}
	\left\lbrace \left( \frac{m }{2\pi} \right)^{\!\frac32}
	T^{\frac{5}{2}} \left[
		\Li{\frac52}\left(
			-e^{\beta(\mu_0 + \frac{H}{2} - \sigma_{\uparrow})}
		\right) +
		\Li{\frac52}\left(
			-e^{\beta(\mu_0 - \frac{H}{2} - \sigma_{\downarrow})}
		\right) 
	\right]
	- \frac{4\pi a}{m} \left[
		(\N-1)N_0^2 - \frac{\Delta N^2}{4}
	\right]
	\right\rbrace.
\eeq
One notes that the dangerous $\N^3$ terms are effectively resummed
into the polylogarithms.

After resumming Hartree-Fock diagrams to all order with the above
procedure, $\Omega_{2a}$ is precisely the next leading correction.
Using equations \eqref{omega2aABSymmetrized} and
\eqref{omega2aSpinSum}, up to fourth overall order,
\beq
\spl{
\frac{\Omega_{2a}}{V} \approx& \;
\frac{k_{\mu}^7 a^2}{12 m \pi^2} \Bigg \lbrace
	\N(\N-1) \left[
		a_{0}^{(2a)} + a_{1}^{(2a)} t^2 + a_{2}^{2a} \frac{\N}{4} h^2
		+ \left(
			a_{5}^{(2a)} \frac{\N}{16}+ a_{6}^{(2a)} \frac{\N^2}{16}
		\right) h^4
		+a_{7}^{(2a)} \frac{\N}{4} t^2 h^2
	\right] \\
	&+ \N \left(\frac{\N}{2}-1\right) f_4(t, 0)
	+ \frac{\N^2}{2} f_4(t, h)
\Bigg \rbrace.
}
\eeq

\end{widetext}

The sum of $\Omega_{HF}$ and $\Omega_{2a}$ gives the desired
approximation to the thermodynamic potential.

The $\Omega_{2a}$ term is of the order $\order{\N^2}$.  For $\N > 2$,
this brings its size closer to the dominating free-gas contribution,
which only scales as $\order{\N}$.  This is the advertised large-$\N$ 
enhancement, and we hope that this will make the quantitative
measurement of the non-analyticity less difficult.

\section{Discussion and Conclusion}

We have presented the equation of state for a SU($\N$) Fermi gas, that 
can in principle be tested in a cold atom experiment setup.  We found 
that the thermodynamic potential $\Omega$ depends non-analytically on 
temperature $T$ and effective magnetic field $\mathbf{H}$, and displays
a crossover behavior as the ratio of $T$ and $\mathbf{H}$ is
continuously varied.  There is a potential enhancement of this
non-analytic behavior if $\N > 2$.

The familiar Ginzburg-Landau (GL) paradigm asserts that, away from a 
phase transition, the thermodynamic behavior of a physical system
should be analytic. This is in direct contrast with our result, where
the equation of state is non-analytic for any non-zero strength of 
interaction.  Yet the qualitative behavior seen in figure 
\ref{phaseDiagram}, even though only at a higher order, is very much 
reminiscent of what is seen near a typical GL critical point.

This result should come as hardly a surprise.  Recall that much of the
GL critical phenomenology rests on one single assertion: a diverging 
correlation length.  The particle-hole pair excitation in our present 
problem exactly fills the role of infinite-ranged correlation, and
this is independent from interaction strength.  In this sense, a FL in 
the normal phase is always ``critical'' in the sub-leading order.

In order to compute the equation of state, we employed a two-step 
approach in this paper.  Analytic terms up to fourth order in $t$ and
$h$ were obtained from considering scattering processes with arbitrary 
momentum transfer in the dilute Fermi gas, and the non-analytic terms
at fourth overall order were then found by considering only
small-momentum scattering processes, with the appropriate asymptotic
form for the particle-hole pair Green's function being used instead
of the exact form.

One is able to do this because, from the Fermi gas picture, it can be 
seen that these ``small-momentum-transfer, on-Fermi-surface''
processes are precisely those contributing to the non-analyticity.
The two steps are thus complimentary, and between them provide the
full answer to the calculation.  This also serves as yet another 
confirmation of the oft-cited observation: ``the leading non-analytic 
correction is universal to all FL'' \cite{Pethick1973}.  

Our analysis is limited to second order perturbation theory.  On one 
hand, cold atom gas experiments have achieved the dilute regime where 
this approximation is justified; on the other hand, this approximation 
allows one to analytically obtain a consistent approximation to the 
equation of state.  So we do not find the limitation too restrictive.

As pointed out by Chubukov and collaborators \cite{Chubukov2006, 
Maslov2009}, one could carry out the same calculation, but with the
fully renormalized quasi-particle dispersion and scattering amplitudes
in the Fermi liquid theory.  The resultant expression remains valid 
beyond the dilute regime, as long as the Fermi liquid picture is 
applicable.  However, away from the dilute regime, one can no longer 
justify the exclusion of other processes.

Experimental confirmation of this non-analytic behavior will be 
challenging, to say the least.  But we hope to see a closure of this
old but interesting problem.

\begin{acknowledgments}

This work is supported by Ministry of Science and Technology of
Taiwan under grant number MOST-104-2112-M-001-006-MY3.  PTH is also 
supported by MOST-106-2811-M-001-002 and MOST-107-2811-M-001-004.
The authors would like to thank Chi-Ho Cheng for some initial 
collaboration and input.

\end{acknowledgments}

\appendix

\section{Sommerfeld Expansion}

\label{sommerfeldAppen}

The original Sommerfeld expansion concerns with integration of a
function $f(\ve)$ of energy $\ve$, weighted by the Fermi function 
$n_F(\ve) = \frac{1}{e^{\beta \ve} + 1}$.  One expands the function 
$f(\ve)$ as a power series of $\ve$, and reduces the original integral 
into the sum of moments of the Fermi function.

In the present work our integration variable is the single-particle 
momentum $\bm{k}$, rather than the energy.  Assuming $f(k)$ depends
only on the magnitude $ k = \vert \bm{k} \vert$, we adopt the original 
procedure into the following:
\beq
\spl{
\int \! \intd^3 \bm{k} \, &\left[
	n_a(k) \; - \Theta \! \left( \! \mu_a - \frac{k^2}{2m} \right) \!
	\right] f(\vert k \vert) \\
&= \sum_{n=0}^{\infty} \tau_{2n+2} \int \! \intd^3 \bm{k} \; 
	\delta \left( \!  k - \sqrt{2 m \mu_a} \right) \\
	& \quad \times \frac{1}{k^2}
		\left[ \left( \frac{m}{k}
			\frac{\partial}{\partial k} \right)^{2n+1}
		k f(k)
	\right],
}
\label{sommerfeldFormula}
\eeq
where $n_a$ is given in \eqref{freeGasNotation}, $\Theta$ is the
Heaviside step function, and the coefficient $\tau_{n}$ is defined
to be
\beq
\tau_{n} = \frac{2m}{(n-1)!} \int_{0}^{\infty} \! \intd \ve
			\, n_F(\ve) \, \ve^{n-a}.
\eeq

Generally, the procedure outlined above does not commute with a
momentum cutoff.  Any cutoff must be implemented formally as a
weighting function multiplied to the original integrand, and the
partial derivatives in \eqref{sommerfeldFormula} should act on the
cutoff function equally.  We implemented a sharp cutoff using a
Heaviside step function to analyze the origin in the momentum space
of non-analyticity in $\Omega_{2a}$.

\section{Crossover Function $\chi$ at Small $h$}

The small-$h$ expansion of $\chi$, given in
\eqref{crossoverChiLargeT}, is quite tricky to derive.  Unlike its
small-$t$ counterpart, the region where $\ve < H_{ab}$ is not
suppressed in the integral.  A straight series expansion in $h$ is
therefore doomed with infrared divergences.

As the expression \eqref{crossoverChi} is manifestly even in $h$, one
can take $h > 0$.  At fixed $t$, we define
\beq
c(\lambda) \equiv \frac{1}{t^4}\tilde{\chi}(t, h).
\eeq
where $\lambda = h/t$.  By construction $c(0)$ vanishes.  And by
being an even function, its first derivative at zero $c'(0)$ also
vanishes.  Our strategy will be to evaluate the second derivative 
$c''(\lambda)$, and then integrate it twice to get back to $c$.

First, the $\ve$-integral in \eqref{crossoverChi} must be
reinterpreted as
\beq
\int_{0}^{\infty} \intd \ve \rightarrow
	\left( \int_{0}^{H^{-}} \intd \ve
		+ \int_{H^{+}}^{\infty} \intd \ve \right).
\eeq
This does not affect $c(h)$ itself in any way, but allows one to 
interchange the order of $\ve$-integration and $h$-differentiation.
Also a convergence factor $x^{-2\epsilon}$ with
$\epsilon \rightarrow 0^{+}$ is needed: even though the end result
is itself finite, we will break the integral into multiple
(diverging) parts, and the convergence factor consistently
regularizes these fictitious divergences.  One may then write:
\beq
c''(\lambda) =
	-\frac{3}{8\pi^2} \int_{0}^{\infty} \! \intd x \;
	\left( \frac{2x^{1-2\epsilon}}{e^x - 1}
		+ \frac{2 \lambda^2}{e^x -1}
		\frac{x^{1-2\epsilon}}{x^2 - \lambda^2} \right).
\label{cIntermediate}
\eeq
The first term in the above integral can be straightforwardly
integrated, yielding $-1/8$.  

\begin{figure}
\includegraphics[scale=0.6]{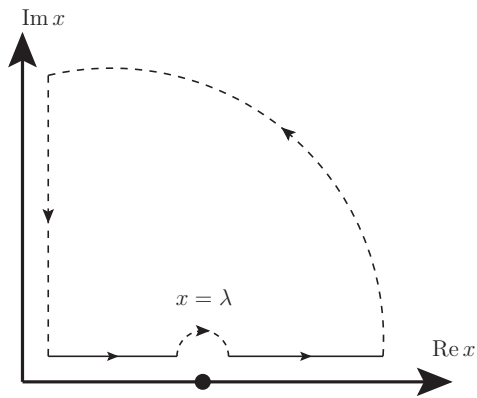}
	\caption{\label{contour} The contour for \eqref{cIntermediate}.
	The entire loop integrate to zero, while the solid part is the
	desired path.}
\end{figure}

For the second piece, one needs to take $x$ to the complex plane, and
the path of integration is as shown in figure \ref{contour}.  One may
complete the contour into a close loop as shown, which integrates to
zero identically.  The arc at infinity vanishes due to the Bose
function.  The required $c''$ is therefore the sum of the small 
semicircle around $x = \lambda$ and the line integral along the 
imaginary-$x$ axis.

The other necessary trick is to rewrite the Bose function as
\beq
\frac{1}{e^{x}-1} =
	-\frac{1}{2} + \frac{1}{x}
	+ \sum_{k = 1}^{\infty} \frac{2 x}{x^2 + (2\pi k)^2}.
\label{boseDecompose}
\eeq
and carry out the integration term by term.

The small semicircle around the $x = \lambda$ pole is easily
evaluated using the residue theorem.  With the aforementioned
convergence factor $x^{-2\epsilon}$, the integral along the
imaginary-$x$ axis and the sum over $k$ can be performed using
standard $\epsilon$-expansion and $\zeta$-function regularization 
techniques, respectively.  Individually the pieces in 
\eqref{boseDecompose} yields $1/\epsilon$ poles, which cancel among 
themselves.  After taking $\epsilon \rightarrow 0$, the end result
is
\beq
\spl{
c''(\lambda) = & \;
	-\frac{3}{8\pi^2} \Bigg[
		\frac{\pi^2}{3} \,
		+ \lambda^2 \ln \vert \lambda \vert
		+ \lambda^2 (\gamma_E - \ln 2\pi) \\
		& + \lambda^2 \sum_{n=1}^{\infty} (-1)^n \zeta(2n+1) \left(
			\frac{\lambda}{2\pi} 
		\right)^{\! 2n}	\Bigg].
}
\eeq

Integrating this result twice with respect to $lambda$, and imposing
the condition $c(0) = c'(0) = 0$, one recovers
\eqref{crossoverChiLargeT} as desired.

\bibliography{fermiLiquid.bib}

\begin{thebibliography}{45}%
\makeatletter
\providecommand \@ifxundefined [1]{%
 \@ifx{#1\undefined}
}%
\providecommand \@ifnum [1]{%
 \ifnum #1\expandafter \@firstoftwo
 \else \expandafter \@secondoftwo
 \fi
}%
\providecommand \@ifx [1]{%
 \ifx #1\expandafter \@firstoftwo
 \else \expandafter \@secondoftwo
 \fi
}%
\providecommand \natexlab [1]{#1}%
\providecommand \enquote  [1]{``#1''}%
\providecommand \bibnamefont  [1]{#1}%
\providecommand \bibfnamefont [1]{#1}%
\providecommand \citenamefont [1]{#1}%
\providecommand \href@noop [0]{\@secondoftwo}%
\providecommand \href [0]{\begingroup \@sanitize@url \@href}%
\providecommand \@href[1]{\@@startlink{#1}\@@href}%
\providecommand \@@href[1]{\endgroup#1\@@endlink}%
\providecommand \@sanitize@url [0]{\catcode `\\12\catcode `\$12\catcode
  `\&12\catcode `\#12\catcode `\^12\catcode `\_12\catcode `\%12\relax}%
\providecommand \@@startlink[1]{}%
\providecommand \@@endlink[0]{}%
\providecommand \url  [0]{\begingroup\@sanitize@url \@url }%
\providecommand \@url [1]{\endgroup\@href {#1}{\urlprefix }}%
\providecommand \urlprefix  [0]{URL }%
\providecommand \Eprint [0]{\href }%
\providecommand \doibase [0]{http://dx.doi.org/}%
\providecommand \selectlanguage [0]{\@gobble}%
\providecommand \bibinfo  [0]{\@secondoftwo}%
\providecommand \bibfield  [0]{\@secondoftwo}%
\providecommand \translation [1]{[#1]}%
\providecommand \BibitemOpen [0]{}%
\providecommand \bibitemStop [0]{}%
\providecommand \bibitemNoStop [0]{.\EOS\space}%
\providecommand \EOS [0]{\spacefactor3000\relax}%
\providecommand \BibitemShut  [1]{\csname bibitem#1\endcsname}%
\let\auto@bib@innerbib\@empty
\bibitem [{\citenamefont {Doniach}\ and\ \citenamefont
  {Engelsberg}(1966)}]{Doniach1966}%
  \BibitemOpen
  \bibfield  {author} {\bibinfo {author} {\bibfnamefont {S.}~\bibnamefont
  {Doniach}}\ and\ \bibinfo {author} {\bibfnamefont {S.}~\bibnamefont
  {Engelsberg}},\ }\href {\doibase 10.1103/PhysRevLett.17.750} {\bibfield
  {journal} {\bibinfo  {journal} {Physical Review Letters}\ }\textbf {\bibinfo
  {volume} {17}},\ \bibinfo {pages} {750} (\bibinfo {year} {1966})}\BibitemShut
  {NoStop}%
\bibitem [{\citenamefont {Amit}\ \emph
  {et~al.}(1968{\natexlab{a}})\citenamefont {Amit}, \citenamefont {Kane},\ and\
  \citenamefont {Wagner}}]{Amit1968}%
  \BibitemOpen
  \bibfield  {author} {\bibinfo {author} {\bibfnamefont {D.~J.}\ \bibnamefont
  {Amit}}, \bibinfo {author} {\bibfnamefont {J.~W.}\ \bibnamefont {Kane}}, \
  and\ \bibinfo {author} {\bibfnamefont {H.}~\bibnamefont {Wagner}},\ }\href
  {\doibase 10.1103/PhysRev.175.326} {\bibfield  {journal} {\bibinfo  {journal}
  {Physical Review}\ }\textbf {\bibinfo {volume} {175}},\ \bibinfo {pages}
  {326} (\bibinfo {year} {1968}{\natexlab{a}})}\BibitemShut {NoStop}%
\bibitem [{\citenamefont {Mota}\ \emph {et~al.}(1969)\citenamefont {Mota},
  \citenamefont {Platzeck}, \citenamefont {Rapp},\ and\ \citenamefont
  {Wheatley}}]{Mota1969}%
  \BibitemOpen
  \bibfield  {author} {\bibinfo {author} {\bibfnamefont {A.~C.}\ \bibnamefont
  {Mota}}, \bibinfo {author} {\bibfnamefont {R.~P.}\ \bibnamefont {Platzeck}},
  \bibinfo {author} {\bibfnamefont {R.}~\bibnamefont {Rapp}}, \ and\ \bibinfo
  {author} {\bibfnamefont {J.~C.}\ \bibnamefont {Wheatley}},\ }\href {\doibase
  10.1103/PhysRev.177.266} {\bibfield  {journal} {\bibinfo  {journal} {Physical
  Review}\ }\textbf {\bibinfo {volume} {177}},\ \bibinfo {pages} {266}
  (\bibinfo {year} {1969})}\BibitemShut {NoStop}%
\bibitem [{\citenamefont {Pethick}\ and\ \citenamefont
  {Carneiro}(1973)}]{Pethick1973}%
  \BibitemOpen
  \bibfield  {author} {\bibinfo {author} {\bibfnamefont {C.~J.}\ \bibnamefont
  {Pethick}}\ and\ \bibinfo {author} {\bibfnamefont {G.~M.}\ \bibnamefont
  {Carneiro}},\ }\href {\doibase 10.1103/PhysRevA.7.304} {\bibfield  {journal}
  {\bibinfo  {journal} {Physical Review A}\ }\textbf {\bibinfo {volume} {7}},\
  \bibinfo {pages} {304} (\bibinfo {year} {1973})}\BibitemShut {NoStop}%
\bibitem [{\citenamefont {Carneiro}\ and\ \citenamefont
  {Pethick}(1975)}]{Carneiro1975}%
  \BibitemOpen
  \bibfield  {author} {\bibinfo {author} {\bibfnamefont {G.~M.}\ \bibnamefont
  {Carneiro}}\ and\ \bibinfo {author} {\bibfnamefont {C.~J.}\ \bibnamefont
  {Pethick}},\ }\href {\doibase 10.1103/PhysRevB.11.1106} {\bibfield  {journal}
  {\bibinfo  {journal} {Physical Review B}\ }\textbf {\bibinfo {volume} {11}},\
  \bibinfo {pages} {1106} (\bibinfo {year} {1975})}\BibitemShut {NoStop}%
\bibitem [{\citenamefont {Carneiro}\ and\ \citenamefont
  {Pethick}(1977)}]{Carneiro1977}%
  \BibitemOpen
  \bibfield  {author} {\bibinfo {author} {\bibfnamefont {G.~M.}\ \bibnamefont
  {Carneiro}}\ and\ \bibinfo {author} {\bibfnamefont {C.~J.}\ \bibnamefont
  {Pethick}},\ }\href {\doibase 10.1103/PhysRevB.16.1933} {\bibfield  {journal}
  {\bibinfo  {journal} {Physical Review B}\ }\textbf {\bibinfo {volume} {16}},\
  \bibinfo {pages} {1933} (\bibinfo {year} {1977})}\BibitemShut {NoStop}%
\bibitem [{\citenamefont {Belitz}\ \emph {et~al.}(1997)\citenamefont {Belitz},
  \citenamefont {Kirkpatrick},\ and\ \citenamefont {Vojta}}]{Belitz1997}%
  \BibitemOpen
  \bibfield  {author} {\bibinfo {author} {\bibfnamefont {D.}~\bibnamefont
  {Belitz}}, \bibinfo {author} {\bibfnamefont {T.~R.}\ \bibnamefont
  {Kirkpatrick}}, \ and\ \bibinfo {author} {\bibfnamefont {T.}~\bibnamefont
  {Vojta}},\ }\href {\doibase 10.1103/PhysRevB.55.9452} {\bibfield  {journal}
  {\bibinfo  {journal} {Physical Review B}\ }\textbf {\bibinfo {volume} {55}},\
  \bibinfo {pages} {9452} (\bibinfo {year} {1997})},\ \Eprint
  {http://arxiv.org/abs/9611099} {arXiv:9611099 [cond-mat]} \BibitemShut
  {NoStop}%
\bibitem [{\citenamefont {Chitov}\ and\ \citenamefont
  {Millis}(2001)}]{Chitov2001}%
  \BibitemOpen
  \bibfield  {author} {\bibinfo {author} {\bibfnamefont {G.~Y.}\ \bibnamefont
  {Chitov}}\ and\ \bibinfo {author} {\bibfnamefont {A.~J.}\ \bibnamefont
  {Millis}},\ }\href {\doibase 10.1103/PhysRevB.64.054414} {\bibfield
  {journal} {\bibinfo  {journal} {Physical Review B}\ }\textbf {\bibinfo
  {volume} {64}},\ \bibinfo {pages} {054414} (\bibinfo {year} {2001})},\
  \Eprint {http://arxiv.org/abs/0103155v1} {arXiv:0103155v1 [arXiv:cond-mat]}
  \BibitemShut {NoStop}%
\bibitem [{\citenamefont {Misawa}(2001)}]{Misawa2001}%
  \BibitemOpen
  \bibfield  {author} {\bibinfo {author} {\bibfnamefont {S.}~\bibnamefont
  {Misawa}},\ }\href {\doibase 10.1016/S0921-4526(00)00597-4} {\bibfield
  {journal} {\bibinfo  {journal} {Physica B: Condensed Matter}\ }\textbf
  {\bibinfo {volume} {294-295}},\ \bibinfo {pages} {10} (\bibinfo {year}
  {2001})}\BibitemShut {NoStop}%
\bibitem [{\citenamefont {Chubukov}\ and\ \citenamefont
  {Maslov}(2003)}]{Chubukov2003}%
  \BibitemOpen
  \bibfield  {author} {\bibinfo {author} {\bibfnamefont {A.~V.}\ \bibnamefont
  {Chubukov}}\ and\ \bibinfo {author} {\bibfnamefont {D.~L.}\ \bibnamefont
  {Maslov}},\ }\href {\doibase 10.1103/PhysRevB.68.155113} {\bibfield
  {journal} {\bibinfo  {journal} {Physical Review B}\ }\textbf {\bibinfo
  {volume} {68}},\ \bibinfo {pages} {155113} (\bibinfo {year} {2003})},\
  \Eprint {http://arxiv.org/abs/0305022} {arXiv:0305022 [cond-mat]}
  \BibitemShut {NoStop}%
\bibitem [{\citenamefont {Betouras}\ \emph {et~al.}(2005)\citenamefont
  {Betouras}, \citenamefont {Efremov},\ and\ \citenamefont
  {Chubukov}}]{Betouras2005}%
  \BibitemOpen
  \bibfield  {author} {\bibinfo {author} {\bibfnamefont {J.}~\bibnamefont
  {Betouras}}, \bibinfo {author} {\bibfnamefont {D.}~\bibnamefont {Efremov}}, \
  and\ \bibinfo {author} {\bibfnamefont {A.}~\bibnamefont {Chubukov}},\ }\href
  {\doibase 10.1103/PhysRevB.72.115112} {\bibfield  {journal} {\bibinfo
  {journal} {Physical Review B}\ }\textbf {\bibinfo {volume} {72}},\ \bibinfo
  {pages} {115112} (\bibinfo {year} {2005})},\ \Eprint
  {http://arxiv.org/abs/0506083} {arXiv:0506083 [cond-mat]} \BibitemShut
  {NoStop}%
\bibitem [{\citenamefont {Chubukov}\ \emph {et~al.}(2005)\citenamefont
  {Chubukov}, \citenamefont {Maslov}, \citenamefont {Gangadharaiah},\ and\
  \citenamefont {Glazman}}]{Chubukov2005}%
  \BibitemOpen
  \bibfield  {author} {\bibinfo {author} {\bibfnamefont {A.~V.}\ \bibnamefont
  {Chubukov}}, \bibinfo {author} {\bibfnamefont {D.~L.}\ \bibnamefont
  {Maslov}}, \bibinfo {author} {\bibfnamefont {S.}~\bibnamefont
  {Gangadharaiah}}, \ and\ \bibinfo {author} {\bibfnamefont {L.~I.}\
  \bibnamefont {Glazman}},\ }\href {\doibase 10.1103/PhysRevLett.95.026402}
  {\bibfield  {journal} {\bibinfo  {journal} {Physical Review Letters}\
  }\textbf {\bibinfo {volume} {95}},\ \bibinfo {pages} {026402} (\bibinfo
  {year} {2005})},\ \Eprint {http://arxiv.org/abs/0502542} {arXiv:0502542
  [cond-mat]} \BibitemShut {NoStop}%
\bibitem [{\citenamefont {Chubukov}\ \emph {et~al.}(2006)\citenamefont
  {Chubukov}, \citenamefont {Maslov},\ and\ \citenamefont
  {Millis}}]{Chubukov2006}%
  \BibitemOpen
  \bibfield  {author} {\bibinfo {author} {\bibfnamefont {A.~V.}\ \bibnamefont
  {Chubukov}}, \bibinfo {author} {\bibfnamefont {D.~L.}\ \bibnamefont
  {Maslov}}, \ and\ \bibinfo {author} {\bibfnamefont {A.~J.}\ \bibnamefont
  {Millis}},\ }\href {\doibase 10.1103/PhysRevB.73.045128} {\bibfield
  {journal} {\bibinfo  {journal} {Physical Review B}\ }\textbf {\bibinfo
  {volume} {73}},\ \bibinfo {pages} {045128} (\bibinfo {year}
  {2006})}\BibitemShut {NoStop}%
\bibitem [{\citenamefont {Maslov}\ and\ \citenamefont
  {Chubukov}(2009)}]{Maslov2009}%
  \BibitemOpen
  \bibfield  {author} {\bibinfo {author} {\bibfnamefont {D.~L.}\ \bibnamefont
  {Maslov}}\ and\ \bibinfo {author} {\bibfnamefont {A.~V.}\ \bibnamefont
  {Chubukov}},\ }\href {\doibase 10.1103/PhysRevB.79.075112} {\bibfield
  {journal} {\bibinfo  {journal} {Physical Review B}\ }\textbf {\bibinfo
  {volume} {79}},\ \bibinfo {pages} {075112} (\bibinfo {year}
  {2009})}\BibitemShut {NoStop}%
\bibitem [{\citenamefont {Abel}\ \emph {et~al.}(1966)\citenamefont {Abel},
  \citenamefont {Anderson}, \citenamefont {Black},\ and\ \citenamefont
  {Wheatley}}]{Abel1966}%
  \BibitemOpen
  \bibfield  {author} {\bibinfo {author} {\bibfnamefont {W.~R.}\ \bibnamefont
  {Abel}}, \bibinfo {author} {\bibfnamefont {A.~C.}\ \bibnamefont {Anderson}},
  \bibinfo {author} {\bibfnamefont {W.~C.}\ \bibnamefont {Black}}, \ and\
  \bibinfo {author} {\bibfnamefont {J.~C.}\ \bibnamefont {Wheatley}},\ }\href
  {\doibase 10.1103/PhysRev.147.111} {\bibfield  {journal} {\bibinfo  {journal}
  {Physical Review}\ }\textbf {\bibinfo {volume} {147}},\ \bibinfo {pages}
  {111} (\bibinfo {year} {1966})}\BibitemShut {NoStop}%
\bibitem [{\citenamefont {Greywall}(1983)}]{Greywall1983}%
  \BibitemOpen
  \bibfield  {author} {\bibinfo {author} {\bibfnamefont {D.~S.}\ \bibnamefont
  {Greywall}},\ }\href {\doibase 10.1103/PhysRevB.27.2747} {\bibfield
  {journal} {\bibinfo  {journal} {Physical Review B}\ }\textbf {\bibinfo
  {volume} {27}},\ \bibinfo {pages} {2747} (\bibinfo {year}
  {1983})}\BibitemShut {NoStop}%
\bibitem [{\citenamefont {Coffey}\ and\ \citenamefont
  {Pethick}(1986)}]{Coffey1986}%
  \BibitemOpen
  \bibfield  {author} {\bibinfo {author} {\bibfnamefont {D.}~\bibnamefont
  {Coffey}}\ and\ \bibinfo {author} {\bibfnamefont {C.~J.}\ \bibnamefont
  {Pethick}},\ }\href {\doibase 10.1103/PhysRevB.33.7508} {\bibfield  {journal}
  {\bibinfo  {journal} {Physical Review B}\ }\textbf {\bibinfo {volume} {33}},\
  \bibinfo {pages} {7508} (\bibinfo {year} {1986})}\BibitemShut {NoStop}%
\bibitem [{\citenamefont {{Van Der Meulen}}\ \emph {et~al.}(1990)\citenamefont
  {{Van Der Meulen}}, \citenamefont {Tarnawski}, \citenamefont {{De Visser}},
  \citenamefont {Franse}, \citenamefont {Perenboom}, \citenamefont {Althof},\
  and\ \citenamefont {{Van Kempen}}}]{VanDerMeulen1990}%
  \BibitemOpen
  \bibfield  {author} {\bibinfo {author} {\bibfnamefont {H.~P.}\ \bibnamefont
  {{Van Der Meulen}}}, \bibinfo {author} {\bibfnamefont {Z.}~\bibnamefont
  {Tarnawski}}, \bibinfo {author} {\bibfnamefont {A.}~\bibnamefont {{De
  Visser}}}, \bibinfo {author} {\bibfnamefont {J.~J.}\ \bibnamefont {Franse}},
  \bibinfo {author} {\bibfnamefont {J.~A.}\ \bibnamefont {Perenboom}}, \bibinfo
  {author} {\bibfnamefont {D.}~\bibnamefont {Althof}}, \ and\ \bibinfo {author}
  {\bibfnamefont {H.}~\bibnamefont {{Van Kempen}}},\ }\href {\doibase
  10.1103/PhysRevB.41.9352} {\bibfield  {journal} {\bibinfo  {journal}
  {Physical Review B}\ }\textbf {\bibinfo {volume} {41}},\ \bibinfo {pages}
  {9352} (\bibinfo {year} {1990})}\BibitemShut {NoStop}%
\bibitem [{\citenamefont {Coffey}\ and\ \citenamefont
  {Pethick}(1988)}]{Coffey1988}%
  \BibitemOpen
  \bibfield  {author} {\bibinfo {author} {\bibfnamefont {D.}~\bibnamefont
  {Coffey}}\ and\ \bibinfo {author} {\bibfnamefont {C.~J.}\ \bibnamefont
  {Pethick}},\ }\href {\doibase 10.1103/PhysRevB.37.442} {\bibfield  {journal}
  {\bibinfo  {journal} {Physical Review B}\ }\textbf {\bibinfo {volume} {37}},\
  \bibinfo {pages} {442} (\bibinfo {year} {1988})}\BibitemShut {NoStop}%
\bibitem [{\citenamefont {Vojta}\ \emph {et~al.}(1997)\citenamefont {Vojta},
  \citenamefont {Belitz}, \citenamefont {Narayanan},\ and\ \citenamefont
  {Kirkpatrick}}]{Vojta1997}%
  \BibitemOpen
  \bibfield  {author} {\bibinfo {author} {\bibfnamefont {T.}~\bibnamefont
  {Vojta}}, \bibinfo {author} {\bibfnamefont {D.}~\bibnamefont {Belitz}},
  \bibinfo {author} {\bibfnamefont {R.}~\bibnamefont {Narayanan}}, \ and\
  \bibinfo {author} {\bibfnamefont {T.}~\bibnamefont {Kirkpatrick}},\ }\href
  {\doibase 10.1007/s002570050399} {\bibfield  {journal} {\bibinfo  {journal}
  {Zeitschrift f{\"{u}}r Physik B Condensed Matter}\ }\textbf {\bibinfo
  {volume} {103}},\ \bibinfo {pages} {451} (\bibinfo {year}
  {1997})}\BibitemShut {NoStop}%
\bibitem [{\citenamefont {Belitz}\ \emph {et~al.}(1999)\citenamefont {Belitz},
  \citenamefont {Kirkpatrick},\ and\ \citenamefont {Vojta}}]{Belitz1999}%
  \BibitemOpen
  \bibfield  {author} {\bibinfo {author} {\bibfnamefont {D.}~\bibnamefont
  {Belitz}}, \bibinfo {author} {\bibfnamefont {T.~R.}\ \bibnamefont
  {Kirkpatrick}}, \ and\ \bibinfo {author} {\bibfnamefont {T.}~\bibnamefont
  {Vojta}},\ }\href {\doibase 10.1103/PhysRevLett.82.4707} {\bibfield
  {journal} {\bibinfo  {journal} {Physical Review Letters}\ }\textbf {\bibinfo
  {volume} {82}},\ \bibinfo {pages} {4707} (\bibinfo {year} {1999})},\ \Eprint
  {http://arxiv.org/abs/9812420} {arXiv:9812420 [cond-mat]} \BibitemShut
  {NoStop}%
\bibitem [{\citenamefont {Pfleiderer}\ \emph {et~al.}(1997)\citenamefont
  {Pfleiderer}, \citenamefont {McMullan}, \citenamefont {Julian},\ and\
  \citenamefont {Lonzarich}}]{Pfleiderer1997}%
  \BibitemOpen
  \bibfield  {author} {\bibinfo {author} {\bibfnamefont {C.}~\bibnamefont
  {Pfleiderer}}, \bibinfo {author} {\bibfnamefont {G.~J.}\ \bibnamefont
  {McMullan}}, \bibinfo {author} {\bibfnamefont {S.~R.}\ \bibnamefont
  {Julian}}, \ and\ \bibinfo {author} {\bibfnamefont {G.~G.}\ \bibnamefont
  {Lonzarich}},\ }\href {\doibase 10.1103/PhysRevB.55.8330} {\bibfield
  {journal} {\bibinfo  {journal} {Physical Review B}\ }\textbf {\bibinfo
  {volume} {55}},\ \bibinfo {pages} {8330} (\bibinfo {year}
  {1997})}\BibitemShut {NoStop}%
\bibitem [{\citenamefont {Amit}\ \emph
  {et~al.}(1968{\natexlab{b}})\citenamefont {Amit}, \citenamefont {Kane},\ and\
  \citenamefont {Wagner}}]{Amit1968a}%
  \BibitemOpen
  \bibfield  {author} {\bibinfo {author} {\bibfnamefont {D.~J.}\ \bibnamefont
  {Amit}}, \bibinfo {author} {\bibfnamefont {J.~W.}\ \bibnamefont {Kane}}, \
  and\ \bibinfo {author} {\bibfnamefont {H.}~\bibnamefont {Wagner}},\ }\href
  {\doibase 10.1103/PhysRev.175.313} {\bibfield  {journal} {\bibinfo  {journal}
  {Physical Review}\ }\textbf {\bibinfo {volume} {175}},\ \bibinfo {pages}
  {313} (\bibinfo {year} {1968}{\natexlab{b}})}\BibitemShut {NoStop}%
\bibitem [{\citenamefont {Dy}\ and\ \citenamefont {Pethick}(1969)}]{Dy1969}%
  \BibitemOpen
  \bibfield  {author} {\bibinfo {author} {\bibfnamefont {K.~S.}\ \bibnamefont
  {Dy}}\ and\ \bibinfo {author} {\bibfnamefont {C.~J.}\ \bibnamefont
  {Pethick}},\ }\href {\doibase 10.1103/PhysRev.185.373} {\bibfield  {journal}
  {\bibinfo  {journal} {Physical Review}\ }\textbf {\bibinfo {volume} {185}},\
  \bibinfo {pages} {373} (\bibinfo {year} {1969})}\BibitemShut {NoStop}%
\bibitem [{\citenamefont {Baym}\ and\ \citenamefont
  {Pethick}(1991)}]{Baym1991}%
  \BibitemOpen
  \bibfield  {author} {\bibinfo {author} {\bibfnamefont {G.}~\bibnamefont
  {Baym}}\ and\ \bibinfo {author} {\bibfnamefont {C.~J.}\ \bibnamefont
  {Pethick}},\ }\href@noop {} {\emph {\bibinfo {title} {{Landau Fermi-Liquid
  Theory: Concepts and Applications}}}}\ (\bibinfo  {publisher} {Wiley-VCH},\
  \bibinfo {year} {1991})\ p.\ \bibinfo {pages} {216}\BibitemShut {NoStop}%
\bibitem [{\citenamefont {Zhang}\ \emph {et~al.}(2009)\citenamefont {Zhang},
  \citenamefont {Su}, \citenamefont {Lu}, \citenamefont {Weng}, \citenamefont
  {Lee},\ and\ \citenamefont {Xiang}}]{Zhang2009}%
  \BibitemOpen
  \bibfield  {author} {\bibinfo {author} {\bibfnamefont {G.~M.}\ \bibnamefont
  {Zhang}}, \bibinfo {author} {\bibfnamefont {Y.~H.}\ \bibnamefont {Su}},
  \bibinfo {author} {\bibfnamefont {Z.~Y.}\ \bibnamefont {Lu}}, \bibinfo
  {author} {\bibfnamefont {Z.~Y.}\ \bibnamefont {Weng}}, \bibinfo {author}
  {\bibfnamefont {D.~H.}\ \bibnamefont {Lee}}, \ and\ \bibinfo {author}
  {\bibfnamefont {T.}~\bibnamefont {Xiang}},\ }\href {\doibase
  10.1209/0295-5075/86/37006} {\bibfield  {journal} {\bibinfo  {journal} {EPL
  (Europhysics Letters)}\ }\textbf {\bibinfo {volume} {86}},\ \bibinfo {pages}
  {37006} (\bibinfo {year} {2009})},\ \Eprint {http://arxiv.org/abs/0809.3874}
  {arXiv:0809.3874} \BibitemShut {NoStop}%
\bibitem [{\citenamefont {Wang}\ \emph {et~al.}(2009)\citenamefont {Wang},
  \citenamefont {Wu}, \citenamefont {Wu}, \citenamefont {Chen}, \citenamefont
  {Xie}, \citenamefont {Ying}, \citenamefont {Yan}, \citenamefont {Liu},\ and\
  \citenamefont {Chen}}]{Wang2009}%
  \BibitemOpen
  \bibfield  {author} {\bibinfo {author} {\bibfnamefont {X.~F.}\ \bibnamefont
  {Wang}}, \bibinfo {author} {\bibfnamefont {T.}~\bibnamefont {Wu}}, \bibinfo
  {author} {\bibfnamefont {G.}~\bibnamefont {Wu}}, \bibinfo {author}
  {\bibfnamefont {H.}~\bibnamefont {Chen}}, \bibinfo {author} {\bibfnamefont
  {Y.~L.}\ \bibnamefont {Xie}}, \bibinfo {author} {\bibfnamefont {J.~J.}\
  \bibnamefont {Ying}}, \bibinfo {author} {\bibfnamefont {Y.~J.}\ \bibnamefont
  {Yan}}, \bibinfo {author} {\bibfnamefont {R.~H.}\ \bibnamefont {Liu}}, \ and\
  \bibinfo {author} {\bibfnamefont {X.~H.}\ \bibnamefont {Chen}},\ }\href
  {\doibase 10.1103/PhysRevLett.102.117005} {\bibfield  {journal} {\bibinfo
  {journal} {Physical Review Letters}\ }\textbf {\bibinfo {volume} {102}},\
  \bibinfo {pages} {100} (\bibinfo {year} {2009})},\ \Eprint
  {http://arxiv.org/abs/0806.2452} {arXiv:0806.2452} \BibitemShut {NoStop}%
\bibitem [{\citenamefont {Klingeler}\ \emph {et~al.}(2010)\citenamefont
  {Klingeler}, \citenamefont {Leps}, \citenamefont {Hellmann}, \citenamefont
  {Popa}, \citenamefont {Stockert}, \citenamefont {Hess}, \citenamefont
  {Kataev}, \citenamefont {Grafe}, \citenamefont {Hammerath}, \citenamefont
  {Lang}, \citenamefont {Wurmehl}, \citenamefont {Behr}, \citenamefont
  {Harnagea}, \citenamefont {Singh},\ and\ \citenamefont
  {B{\"{u}}chner}}]{Klingeler2010}%
  \BibitemOpen
  \bibfield  {author} {\bibinfo {author} {\bibfnamefont {R.}~\bibnamefont
  {Klingeler}}, \bibinfo {author} {\bibfnamefont {N.}~\bibnamefont {Leps}},
  \bibinfo {author} {\bibfnamefont {I.}~\bibnamefont {Hellmann}}, \bibinfo
  {author} {\bibfnamefont {A.}~\bibnamefont {Popa}}, \bibinfo {author}
  {\bibfnamefont {U.}~\bibnamefont {Stockert}}, \bibinfo {author}
  {\bibfnamefont {C.}~\bibnamefont {Hess}}, \bibinfo {author} {\bibfnamefont
  {V.}~\bibnamefont {Kataev}}, \bibinfo {author} {\bibfnamefont {H.-J.}\
  \bibnamefont {Grafe}}, \bibinfo {author} {\bibfnamefont {F.}~\bibnamefont
  {Hammerath}}, \bibinfo {author} {\bibfnamefont {G.}~\bibnamefont {Lang}},
  \bibinfo {author} {\bibfnamefont {S.}~\bibnamefont {Wurmehl}}, \bibinfo
  {author} {\bibfnamefont {G.}~\bibnamefont {Behr}}, \bibinfo {author}
  {\bibfnamefont {L.}~\bibnamefont {Harnagea}}, \bibinfo {author}
  {\bibfnamefont {S.}~\bibnamefont {Singh}}, \ and\ \bibinfo {author}
  {\bibfnamefont {B.}~\bibnamefont {B{\"{u}}chner}},\ }\href {\doibase
  10.1103/PhysRevB.81.024506} {\bibfield  {journal} {\bibinfo  {journal}
  {Physical Review B}\ }\textbf {\bibinfo {volume} {81}},\ \bibinfo {pages}
  {024506} (\bibinfo {year} {2010})},\ \Eprint {http://arxiv.org/abs/0808.0708}
  {arXiv:0808.0708} \BibitemShut {NoStop}%
\bibitem [{\citenamefont {Korshunov}\ \emph {et~al.}(2009)\citenamefont
  {Korshunov}, \citenamefont {Eremin}, \citenamefont {Efremov}, \citenamefont
  {Maslov},\ and\ \citenamefont {Chubukov}}]{Korshunov2009}%
  \BibitemOpen
  \bibfield  {author} {\bibinfo {author} {\bibfnamefont {M.~M.}\ \bibnamefont
  {Korshunov}}, \bibinfo {author} {\bibfnamefont {I.}~\bibnamefont {Eremin}},
  \bibinfo {author} {\bibfnamefont {D.~V.}\ \bibnamefont {Efremov}}, \bibinfo
  {author} {\bibfnamefont {D.~L.}\ \bibnamefont {Maslov}}, \ and\ \bibinfo
  {author} {\bibfnamefont {A.~V.}\ \bibnamefont {Chubukov}},\ }\href {\doibase
  10.1103/PhysRevLett.102.236403} {\bibfield  {journal} {\bibinfo  {journal}
  {Physical Review Letters}\ }\textbf {\bibinfo {volume} {102}},\ \bibinfo
  {pages} {1} (\bibinfo {year} {2009})},\ \Eprint
  {http://arxiv.org/abs/0901.0238} {arXiv:0901.0238} \BibitemShut {NoStop}%
\bibitem [{\citenamefont {Inouye}\ \emph {et~al.}(1998)\citenamefont {Inouye},
  \citenamefont {Andrews}, \citenamefont {Stenger}, \citenamefont {Miesner},
  \citenamefont {Stamper-Kurn},\ and\ \citenamefont {Ketterle}}]{Inouye1998}%
  \BibitemOpen
  \bibfield  {author} {\bibinfo {author} {\bibfnamefont {S.}~\bibnamefont
  {Inouye}}, \bibinfo {author} {\bibfnamefont {M.~R.}\ \bibnamefont {Andrews}},
  \bibinfo {author} {\bibfnamefont {J.}~\bibnamefont {Stenger}}, \bibinfo
  {author} {\bibfnamefont {H.-J.}\ \bibnamefont {Miesner}}, \bibinfo {author}
  {\bibfnamefont {D.~M.}\ \bibnamefont {Stamper-Kurn}}, \ and\ \bibinfo
  {author} {\bibfnamefont {W.}~\bibnamefont {Ketterle}},\ }\href {\doibase
  10.1038/32354} {\bibfield  {journal} {\bibinfo  {journal} {Nature}\ }\textbf
  {\bibinfo {volume} {392}},\ \bibinfo {pages} {151} (\bibinfo {year}
  {1998})}\BibitemShut {NoStop}%
\bibitem [{\citenamefont {Chin}\ \emph {et~al.}(2010)\citenamefont {Chin},
  \citenamefont {Grimm}, \citenamefont {Julienne},\ and\ \citenamefont
  {Tiesinga}}]{Chin2010}%
  \BibitemOpen
  \bibfield  {author} {\bibinfo {author} {\bibfnamefont {C.}~\bibnamefont
  {Chin}}, \bibinfo {author} {\bibfnamefont {R.}~\bibnamefont {Grimm}},
  \bibinfo {author} {\bibfnamefont {P.}~\bibnamefont {Julienne}}, \ and\
  \bibinfo {author} {\bibfnamefont {E.}~\bibnamefont {Tiesinga}},\ }\href
  {\doibase 10.1103/RevModPhys.82.1225} {\bibfield  {journal} {\bibinfo
  {journal} {Reviews of Modern Physics}\ }\textbf {\bibinfo {volume} {82}},\
  \bibinfo {pages} {1225} (\bibinfo {year} {2010})},\ \Eprint
  {http://arxiv.org/abs/0812.1496} {arXiv:0812.1496} \BibitemShut {NoStop}%
\bibitem [{\citenamefont {Cheng}\ and\ \citenamefont {Yip}(2007)}]{Cheng2007}%
  \BibitemOpen
  \bibfield  {author} {\bibinfo {author} {\bibfnamefont {C.-H.}\ \bibnamefont
  {Cheng}}\ and\ \bibinfo {author} {\bibfnamefont {S.-K.}\ \bibnamefont
  {Yip}},\ }\href {\doibase 10.1103/PhysRevB.75.014526} {\bibfield  {journal}
  {\bibinfo  {journal} {Physical Review B}\ }\textbf {\bibinfo {volume} {75}},\
  \bibinfo {pages} {014526} (\bibinfo {year} {2007})},\ \Eprint
  {http://arxiv.org/abs/0611578} {arXiv:0611578 [cond-mat]} \BibitemShut
  {NoStop}%
\bibitem [{\citenamefont {Navon}\ \emph {et~al.}(2010)\citenamefont {Navon},
  \citenamefont {Nascimbene}, \citenamefont {Chevy},\ and\ \citenamefont
  {Salomon}}]{Navon2010}%
  \BibitemOpen
  \bibfield  {author} {\bibinfo {author} {\bibfnamefont {N.}~\bibnamefont
  {Navon}}, \bibinfo {author} {\bibfnamefont {S.}~\bibnamefont {Nascimbene}},
  \bibinfo {author} {\bibfnamefont {F.}~\bibnamefont {Chevy}}, \ and\ \bibinfo
  {author} {\bibfnamefont {C.}~\bibnamefont {Salomon}},\ }\href {\doibase
  10.1126/science.1187582} {\bibfield  {journal} {\bibinfo  {journal}
  {Science}\ }\textbf {\bibinfo {volume} {328}},\ \bibinfo {pages} {729}
  (\bibinfo {year} {2010})},\ \Eprint {http://arxiv.org/abs/1004.1465}
  {arXiv:1004.1465} \BibitemShut {NoStop}%
\bibitem [{\citenamefont {Ku}\ \emph {et~al.}(2012)\citenamefont {Ku},
  \citenamefont {Sommer}, \citenamefont {Cheuk},\ and\ \citenamefont
  {Zwierlein}}]{Ku2012}%
  \BibitemOpen
  \bibfield  {author} {\bibinfo {author} {\bibfnamefont {M.~J.~H.}\
  \bibnamefont {Ku}}, \bibinfo {author} {\bibfnamefont {A.~T.}\ \bibnamefont
  {Sommer}}, \bibinfo {author} {\bibfnamefont {L.~W.}\ \bibnamefont {Cheuk}}, \
  and\ \bibinfo {author} {\bibfnamefont {M.~W.}\ \bibnamefont {Zwierlein}},\
  }\href {\doibase 10.1126/science.1214987} {\bibfield  {journal} {\bibinfo
  {journal} {Science}\ }\textbf {\bibinfo {volume} {335}},\ \bibinfo {pages}
  {563} (\bibinfo {year} {2012})},\ \Eprint {http://arxiv.org/abs/1110.3309}
  {arXiv:1110.3309} \BibitemShut {NoStop}%
\bibitem [{\citenamefont {{Van Houcke}}\ \emph {et~al.}(2012)\citenamefont
  {{Van Houcke}}, \citenamefont {Werner}, \citenamefont {Kozik}, \citenamefont
  {Prokof'ev}, \citenamefont {Svistunov}, \citenamefont {Ku}, \citenamefont
  {Sommer}, \citenamefont {Cheuk}, \citenamefont {Schirotzek},\ and\
  \citenamefont {Zwierlein}}]{VanHoucke2012}%
  \BibitemOpen
  \bibfield  {author} {\bibinfo {author} {\bibfnamefont {K.}~\bibnamefont {{Van
  Houcke}}}, \bibinfo {author} {\bibfnamefont {F.}~\bibnamefont {Werner}},
  \bibinfo {author} {\bibfnamefont {E.}~\bibnamefont {Kozik}}, \bibinfo
  {author} {\bibfnamefont {N.}~\bibnamefont {Prokof'ev}}, \bibinfo {author}
  {\bibfnamefont {B.}~\bibnamefont {Svistunov}}, \bibinfo {author}
  {\bibfnamefont {M.~J.~H.}\ \bibnamefont {Ku}}, \bibinfo {author}
  {\bibfnamefont {A.~T.}\ \bibnamefont {Sommer}}, \bibinfo {author}
  {\bibfnamefont {L.~W.}\ \bibnamefont {Cheuk}}, \bibinfo {author}
  {\bibfnamefont {A.}~\bibnamefont {Schirotzek}}, \ and\ \bibinfo {author}
  {\bibfnamefont {M.~W.}\ \bibnamefont {Zwierlein}},\ }\href {\doibase
  10.1038/nphys2273} {\bibfield  {journal} {\bibinfo  {journal} {Nature
  Physics}\ }\textbf {\bibinfo {volume} {8}},\ \bibinfo {pages} {366} (\bibinfo
  {year} {2012})},\ \Eprint {http://arxiv.org/abs/arXiv:1110.3747v2}
  {arXiv:arXiv:1110.3747v2} \BibitemShut {NoStop}%
\bibitem [{\citenamefont {Desbuquois}\ \emph {et~al.}(2014)\citenamefont
  {Desbuquois}, \citenamefont {Yefsah}, \citenamefont {Chomaz}, \citenamefont
  {Weitenberg}, \citenamefont {Corman}, \citenamefont {Nascimb{\`{e}}ne},\ and\
  \citenamefont {Dalibard}}]{Desbuquois2014}%
  \BibitemOpen
  \bibfield  {author} {\bibinfo {author} {\bibfnamefont {R.}~\bibnamefont
  {Desbuquois}}, \bibinfo {author} {\bibfnamefont {T.}~\bibnamefont {Yefsah}},
  \bibinfo {author} {\bibfnamefont {L.}~\bibnamefont {Chomaz}}, \bibinfo
  {author} {\bibfnamefont {C.}~\bibnamefont {Weitenberg}}, \bibinfo {author}
  {\bibfnamefont {L.}~\bibnamefont {Corman}}, \bibinfo {author} {\bibfnamefont
  {S.}~\bibnamefont {Nascimb{\`{e}}ne}}, \ and\ \bibinfo {author}
  {\bibfnamefont {J.}~\bibnamefont {Dalibard}},\ }\href {\doibase
  10.1103/PhysRevLett.113.020404} {\bibfield  {journal} {\bibinfo  {journal}
  {Physical Review Letters}\ }\textbf {\bibinfo {volume} {113}},\ \bibinfo
  {pages} {020404} (\bibinfo {year} {2014})},\ \Eprint
  {http://arxiv.org/abs/1403.4030} {arXiv:1403.4030} \BibitemShut {NoStop}%
\bibitem [{\citenamefont {Fukuhara}\ \emph {et~al.}(2007)\citenamefont
  {Fukuhara}, \citenamefont {Takasu}, \citenamefont {Kumakura},\ and\
  \citenamefont {Takahashi}}]{Fukuhara2007}%
  \BibitemOpen
  \bibfield  {author} {\bibinfo {author} {\bibfnamefont {T.}~\bibnamefont
  {Fukuhara}}, \bibinfo {author} {\bibfnamefont {Y.}~\bibnamefont {Takasu}},
  \bibinfo {author} {\bibfnamefont {M.}~\bibnamefont {Kumakura}}, \ and\
  \bibinfo {author} {\bibfnamefont {Y.}~\bibnamefont {Takahashi}},\ }\href
  {\doibase 10.1103/PhysRevLett.98.030401} {\bibfield  {journal} {\bibinfo
  {journal} {Physical Review Letters}\ }\textbf {\bibinfo {volume} {98}},\
  \bibinfo {pages} {030401} (\bibinfo {year} {2007})},\ \Eprint
  {http://arxiv.org/abs/0607228} {arXiv:0607228 [arXiv:cond-mat]} \BibitemShut
  {NoStop}%
\bibitem [{\citenamefont {Cazalilla}\ \emph {et~al.}(2009)\citenamefont
  {Cazalilla}, \citenamefont {Ho},\ and\ \citenamefont {Ueda}}]{Cazalilla2009}%
  \BibitemOpen
  \bibfield  {author} {\bibinfo {author} {\bibfnamefont {M.~A.}\ \bibnamefont
  {Cazalilla}}, \bibinfo {author} {\bibfnamefont {A.~F.}\ \bibnamefont {Ho}}, \
  and\ \bibinfo {author} {\bibfnamefont {M.}~\bibnamefont {Ueda}},\ }\href
  {\doibase 10.1088/1367-2630/11/10/103033} {\bibfield  {journal} {\bibinfo
  {journal} {New Journal of Physics}\ }\textbf {\bibinfo {volume} {11}},\
  \bibinfo {pages} {103033} (\bibinfo {year} {2009})},\ \Eprint
  {http://arxiv.org/abs/0905.4948} {arXiv:0905.4948} \BibitemShut {NoStop}%
\bibitem [{\citenamefont {DeSalvo}\ \emph {et~al.}(2010)\citenamefont
  {DeSalvo}, \citenamefont {Yan}, \citenamefont {Mickelson}, \citenamefont
  {{Martinez de Escobar}},\ and\ \citenamefont {Killian}}]{DeSalvo2010}%
  \BibitemOpen
  \bibfield  {author} {\bibinfo {author} {\bibfnamefont {B.~J.}\ \bibnamefont
  {DeSalvo}}, \bibinfo {author} {\bibfnamefont {M.}~\bibnamefont {Yan}},
  \bibinfo {author} {\bibfnamefont {P.~G.}\ \bibnamefont {Mickelson}}, \bibinfo
  {author} {\bibfnamefont {Y.~N.}\ \bibnamefont {{Martinez de Escobar}}}, \
  and\ \bibinfo {author} {\bibfnamefont {T.~C.}\ \bibnamefont {Killian}},\
  }\href {\doibase 10.1103/PhysRevLett.105.030402} {\bibfield  {journal}
  {\bibinfo  {journal} {Physical Review Letters}\ }\textbf {\bibinfo {volume}
  {105}},\ \bibinfo {pages} {030402} (\bibinfo {year} {2010})},\ \Eprint
  {http://arxiv.org/abs/1005.0668} {arXiv:1005.0668} \BibitemShut {NoStop}%
\bibitem [{\citenamefont {Cazalilla}\ and\ \citenamefont
  {Rey}(2014)}]{Cazalilla2014}%
  \BibitemOpen
  \bibfield  {author} {\bibinfo {author} {\bibfnamefont {M.~A.}\ \bibnamefont
  {Cazalilla}}\ and\ \bibinfo {author} {\bibfnamefont {A.~M.}\ \bibnamefont
  {Rey}},\ }\href {\doibase 10.1088/0034-4885/77/12/124401} {\bibfield
  {journal} {\bibinfo  {journal} {Reports on Progress in Physics}\ }\textbf
  {\bibinfo {volume} {77}},\ \bibinfo {pages} {124401} (\bibinfo {year}
  {2014})},\ \Eprint {http://arxiv.org/abs/arXiv:1403.2792v1}
  {arXiv:arXiv:1403.2792v1} \BibitemShut {NoStop}%
\bibitem [{\citenamefont {Cheng}\ and\ \citenamefont {Yip}(2017)}]{Cheng2017}%
  \BibitemOpen
  \bibfield  {author} {\bibinfo {author} {\bibfnamefont {C.-H.}\ \bibnamefont
  {Cheng}}\ and\ \bibinfo {author} {\bibfnamefont {S.-K.}\ \bibnamefont
  {Yip}},\ }\href {\doibase 10.1103/PhysRevA.95.033619} {\bibfield  {journal}
  {\bibinfo  {journal} {Physical Review A}\ }\textbf {\bibinfo {volume} {95}},\
  \bibinfo {pages} {033619} (\bibinfo {year} {2017})},\ \Eprint
  {http://arxiv.org/abs/1612.07886} {arXiv:1612.07886} \BibitemShut {NoStop}%
\bibitem [{\citenamefont {Belitz}\ and\ \citenamefont
  {Vojta}(2005)}]{Belitz2005}%
  \BibitemOpen
  \bibfield  {author} {\bibinfo {author} {\bibfnamefont {D.}~\bibnamefont
  {Belitz}}\ and\ \bibinfo {author} {\bibfnamefont {T.}~\bibnamefont {Vojta}},\
  }\href {\doibase 10.1103/RevModPhys.77.579} {\bibfield  {journal} {\bibinfo
  {journal} {Reviews of Modern Physics}\ }\textbf {\bibinfo {volume} {77}},\
  \bibinfo {pages} {579} (\bibinfo {year} {2005})}\BibitemShut {NoStop}%
\bibitem [{Note1()}]{Note1}%
  \BibitemOpen
  \bibinfo {note} {For the general case of $\protect \mathcal {N}_c> 2$, third
  order terms in magnetic field is possible; see section \ref {largeN}. But it
  is also determined exclusively by FL parameters.}\BibitemShut {Stop}%
\bibitem [{\citenamefont {Kanno}(1970)}]{Kanno1970}%
  \BibitemOpen
  \bibfield  {author} {\bibinfo {author} {\bibfnamefont {S.}~\bibnamefont
  {Kanno}},\ }\href {\doibase 10.1143/PTP.44.813} {\bibfield  {journal}
  {\bibinfo  {journal} {Progress of Theoretical Physics}\ }\textbf {\bibinfo
  {volume} {44}},\ \bibinfo {pages} {813} (\bibinfo {year} {1970})}\BibitemShut
  {NoStop}%
\bibitem [{\citenamefont {Luttinger}\ and\ \citenamefont
  {Ward}(1960)}]{Luttinger1960}%
  \BibitemOpen
  \bibfield  {author} {\bibinfo {author} {\bibfnamefont {J.~M.}\ \bibnamefont
  {Luttinger}}\ and\ \bibinfo {author} {\bibfnamefont {J.~C.}\ \bibnamefont
  {Ward}},\ }\href {\doibase 10.1103/PhysRev.118.1417} {\bibfield  {journal}
  {\bibinfo  {journal} {Physical Review}\ }\textbf {\bibinfo {volume} {118}},\
  \bibinfo {pages} {1417} (\bibinfo {year} {1960})}\BibitemShut {NoStop}%
\end{thebibliography}%

\end{document}